\documentclass[12pt]{article}
\usepackage{amsmath}
\usepackage{amssymb}
\usepackage{slashed}
\usepackage{graphicx}
\usepackage{subfigure}
\usepackage{cite}
\usepackage{url}
\usepackage[small]{caption}
\begin{document}
\baselineskip 0.6cm

\def\bra#1{\langle #1 |}
\def\ket#1{| #1 \rangle}
\def\inner#1#2{\langle #1 | #2 \rangle}

\begin{titlepage}

\begin{flushright}
UCB-PTH-14/41\\
\end{flushright}

\vskip 1.5cm

\begin{center}
{\Large \bf Relativeness in Quantum Gravity:\\
 Limitations and Frame Dependence of Semiclassical Descriptions}

\vskip 0.7cm

{\large Yasunori Nomura, Fabio Sanches, and Sean J. Weinberg}

\vskip 0.4cm

{\it Berkeley Center for Theoretical Physics, Department of Physics,\\
 University of California, Berkeley, CA 94720, USA}

\vskip 0.1cm

{\it Theoretical Physics Group, Lawrence Berkeley National Laboratory,
 CA 94720, USA}

\vskip 0.8cm

\abstract{Consistency between quantum mechanical and general relativistic 
 views of the world is a longstanding problem, which becomes particularly 
 prominent in black hole physics.  We develop a coherent picture addressing 
 this issue by studying the quantum mechanics of an evolving black 
 hole.  After interpreting the Bekenstein-Hawking entropy as the entropy 
 representing the degrees of freedom that are coarse-grained to obtain 
 a semiclassical description from the microscopic theory of quantum 
 gravity, we discuss the properties these degrees of freedom exhibit 
 when viewed from the semiclassical standpoint.  We are led to the 
 conclusion that they show features which we call {\it extreme relativeness} 
 and {\it spacetime-matter duality}---a nontrivial reference frame 
 dependence of their spacetime distribution and the dual roles they 
 play as the ``constituents'' of spacetime and as thermal radiation. 
 We describe black hole formation and evaporation processes in distant 
 and infalling reference frames, showing that these two properties allow 
 us to avoid the arguments for firewalls and to make the existence of 
 the black hole interior consistent with unitary evolution in the sense 
 of complementarity.  Our analysis provides a concrete answer to how 
 information can be preserved at the quantum level throughout the 
 evolution of a black hole, and gives a basic picture of how general 
 coordinate transformations may work at the level of full quantum 
 gravity beyond the approximation of semiclassical theory.}

\end{center}
\end{titlepage}

\section{Introduction}
\label{sec:intro}

In the past decades, it has become increasingly apparent that the concept 
of spacetime must receive substantial revisions when it is treated in 
a fully quantum mechanical manner.  The first clear sign of this came 
from the study of black hole physics~\cite{Preskill:1992tc}.  Consider 
describing a process in which an object falls into a black hole, which 
eventually evaporates, from the viewpoint of a distant observer.  Unitarity 
of quantum mechanics suggests that the information content of the object 
will first be stored in the black hole system, and then emitted back 
to distant space in the form of Hawking radiation~\cite{'tHooft:1990fr}. 
On the other hand, the equivalence principle implies that the object 
should not find anything special at the horizon, when the process is 
described by an observer falling with the object.  These two pictures 
lead to inconsistency if we adopt the standard formulation of quantum 
field theory on curved spacetime, since it allows us to employ a class 
of equal time hypersurfaces (called nice slices) that pass through both 
the fallen object and late Hawking radiation, leading to violation of 
the no-cloning theorem of quantum mechanics~\cite{Wootters:1982zz}.

In the early 90's, a remarkable suggestion to avoid this difficulty---called 
complementarity---was made~\cite{Susskind:1993if}:\ the apparent cloning 
of the information occurring in black hole physics implies that the 
internal spacetime and horizon/Hawking radiation degrees of freedom 
appearing in different, i.e.\ infalling and distant, descriptions are 
not independent.  This signals a breakdown of the naive global spacetime 
picture of general relativity at the quantum level, and it forces us to 
develop a new view of how classical spacetime arises in the full theory 
of quantum gravity.  One of the main purposes of this paper is to present 
a coherent picture of this issue.  We discuss how a series of well-motivated 
hypotheses leads to a consistent view of the effective emergence of global 
spacetime from a fundamental theory of quantum gravity.  In particular, 
we elucidate how this picture avoids the recently raised firewall 
paradox~\cite{Almheiri:2012rt,Almheiri:2013hfa,Marolf:2013dba}, 
which can be viewed as a refined version of the old information 
paradox~\cite{Hawking:1976ra}.  Our analysis provides a concrete 
answer to how the information can be preserved at the quantum level 
in the black hole formation and evaporation processes.

A key element in developing our picture is to identify the origin and nature 
of the ``entropy of spacetime,'' first discovered by Bekenstein and Hawking 
in studying black hole physics~\cite{Bekenstein:1973ur,Hawking:1974sw}. 
In a previous paper~\cite{Nomura:2014yka}, two of us argued that this 
entropy---the Bekenstein-Hawking entropy---is associated with the degrees 
of freedom that are coarse-grained {\it to obtain} the semiclassical 
description of the system:\ quantum theory of matter and radiation on 
a fixed spacetime background.  This picture is consonant with the fact 
that in quantum mechanics, having a well-defined geometry of spacetime, 
e.g.\ a black hole in a well-defined spacetime location, requires taking 
a superposition of an enormous number of energy-momentum eigenstates, 
so we expect that there are many different ways to arrive at the same 
background for the semiclassical theory within the precision allowed 
by quantum mechanics.  This implies that, when a system with a black 
hole is described in a distant reference frame, the information about 
the microstate of the black hole is {\it delocalized} over a large 
spatial region, since it is encoded globally in the way of taking 
the energy-momentum superposition to arrive at the geometry under 
consideration.  In particular, we may naturally identify the spatial 
distribution of this information as that of the gravitational thermal 
entropy calculated using the semiclassical theory.  This leads to 
a fascinating picture:\ the degrees of freedom represented by the 
Bekenstein-Hawking entropy play dual roles of spacetime and matter---they 
represent how the semiclassical geometry is obtained at the microscopic 
level and at the same time can be viewed as the origin of the thermal 
entropy, which is traditionally associated with thermal radiation in 
the semiclassical theory.

The delocalization of the microscopic information described above 
plays an important role in addressing the firewall/information paradox. 
As described in a distant reference frame, a general black hole 
state is specified by the following three classes of indices at 
the microscopic level:
\begin{itemize}
\item
Indices labeling the (field or string theoretic) degrees of freedom 
in the exterior spacetime region, excited over the vacuum of the 
semiclassical theory;%
\footnote{Note that the concepts of the breakdown of a semiclassical 
 description and that of semiclassical {\it field} theory are not the 
 same---there can be phase space regions in which an object can be well 
 described as a string (or brane) propagating in spacetime, but not as 
 a particle.}
\item
Indices labeling the excitations of the stretched horizon;%
\footnote{The stretched horizon is located at a microscopic distance 
 outside of the mathematical horizon, and is regarded as a physical 
 (timelike) membrane which may be physically excited~\cite{Susskind:1993if}.}
\item
Indices representing the degrees of freedom that are coarse-grained 
to obtain the semiclassical description, which we will collectively 
denote by $k$.  The information in $k$ represents how the black hole 
geometry is obtained at the microscopic level, and cannot be resolved 
by semiclassical operators.  It is regarded as being delocalized 
following the spatial distribution of the gravitational thermal entropy, 
calculated using the semiclassical theory.
\end{itemize}
In a distant reference frame, an object falling into the black hole 
is initially described by the first class of indices, and then by the 
second when it hits the stretched horizon.  The information about the 
fallen object will then reside there for, at least, time of order $M 
l_{\rm P}^2 \ln (M l_{\rm P})$ (the scrambling time~\cite{Hayden:2007cs}), 
after which it will be transmitted to the index $k$.  Here, $M$ and 
$l_{\rm P}$ are the mass of the black hole and the Planck length, 
respectively.  Finally, the information in $k$, which is delocalized 
in the whole zone region, will leave the black hole system through 
the Hawking emission, or black hole mining, process.

Since the microscopic information about the black hole is considered 
to be delocalized from the semiclassical standpoint, the Hawking 
emission, or black hole mining, process can be viewed as occurring 
at a {\it macroscopic} distance away from the stretched horizon without 
contradicting information conservation.  In this region, degrees of freedom 
represented by the index $k$ are converted into modes that have clear 
identities as semiclassical excitations, i.e.\ matter or radiation, above 
the spacetime background.  This conversion process, i.e.\ the emission 
of Hawking quanta or the excitation of a mining apparatus, is accompanied 
by the appearance of negative energy excitations, which have {\it negative 
entropies} and propagate inward to the stretched horizon.  As we will 
see, the microscopic dynamics of quantum gravity allows these processes 
to occur unitarily without violating causality among events described 
in low energy quantum field theory.  This picture avoids firewalls as 
well as information cloning.

In the description based on a distant reference frame, a falling object 
can be described by the semiclassical theory only until it hits the 
stretched horizon, after which it goes outside the applicability domain 
of the theory.  We may, however, describe the fate of the object using 
the semiclassical language somewhat longer by performing a {\it reference 
frame change}, specifically until the object hits a singularity, after 
which there is no reference frame that admits a semiclassical description 
of the object.  This reference frame change is the heart of complementarity:\ 
the emergence of global spacetime in the classical limit.  We argue 
that while descriptions in different reference frames (the descriptions 
before and after a complementarity transformation) apparently look 
very different, e.g.\ in locations of the degrees of freedom representing 
the microscopic information of the black hole, their predictions 
about the same physical question are consistent with each other. 
This consistency is ensured by an intricate interplay between 
the properties of microscopic information and the causal structure 
of spacetime.

It is striking that the concept of spacetime, e.g.\ the region in 
which a semiclassical description is applicable, depends on a reference 
frame.  This extreme ``relativeness'' of the description is a result 
of nonzero Newton's constant $G_{\rm N}$.  The situation is analogous 
to what happened when the speed of light, $c$, was realized to be 
finite~\cite{Nomura:2011rb}:\ in Galilean physics ($c = \infty$) 
a change of the reference frame leads only to a constant shift of all 
the velocities, while in special relativity ($c = \mbox{finite}$) it 
also alters temporal and spatial lengths (time dilation and Lorentz 
contraction) and makes the concept of simultaneity relative.  With 
gravity ($G_{\rm N} \neq 0$), even the concept of spacetime becomes 
relative.  The trend is consistent---as we ``turn on'' fundamental 
constants in nature ($c = \infty \rightarrow \mbox{finite}$ and 
$G_{\rm N} = 0 \rightarrow\,\, \neq 0$), physical descriptions become 
more and more relative:\ descriptions of the same physical system in 
different reference frames appear to differ more and more.

The organization of this paper is the following.  In 
Section~\ref{sec:failure}, we discuss some basic aspects of 
the breakdown of global spacetime, setting up the stage for later 
discussions.  In Sections~\ref{sec:distant} and \ref{sec:infalling}, 
we describe how our picture addresses the problem of black hole 
formation and evaporation.  We discuss the quantum structure of 
black hole microstates and the unitary flow of information as viewed 
from a distant reference frame (in Section~\ref{sec:distant}), and 
how it can be consistent with the existence of interior spacetime 
(in Section~\ref{sec:infalling}).  In particular, we elucidate 
how this picture addresses the arguments for firewalls and provides 
a consistent resolution to the black hole information paradox.  
In Section~\ref{sec:summary}, we give our summary by presenting 
a grand picture of the structure of quantum gravity implied by 
our analysis of a system with a black hole.

Throughout the paper, we adopt the Schr\"{o}dinger picture for quantum 
evolution, and use natural units in which $\hbar = c = 1$ unless otherwise 
stated.  We limit our discussions to 4-dimensional spacetime, although 
we do not expect difficulty in extending to other dimensions.  The value 
of the Planck length in our universe is $l_{\rm P} = G_{\rm N}^{1/2} 
\simeq 1.62 \times 10^{-35}~{\rm m}$.  A concise summary of the 
implications of our framework for black hole physics can be found 
in Ref.~\cite{Nomura:2014woa}.

\section{Failure of Global Spacetime}
\label{sec:failure}

As described in the introduction, semiclassical theory applied to an 
entire global spacetime leads to overcounting of the true degrees of 
freedom at the quantum level.  This implies that in the full theory of 
quantum gravity, a semiclassical description of physics emerges only in 
some limited sense.  Here we discuss basic aspects of this limitation, 
setting up the stage for later discussions.

The idea of complementarity~\cite{Susskind:1993if} is that the overcounting 
inherent in the global spacetime picture may be avoided if we limit 
our description to what a single ``observer''---represented by a single 
worldline in spacetime---can causally access.  Depending on which observer 
we choose, we obtain different descriptions of the system, which are 
supposed to be equivalent.  Since the events an observer can see lie 
within the causal patch associated with the worldline representing the 
observer, we may assume that this causal patch is the spacetime region 
a single such description may represent.  In particular, one may postulate 
the following~\cite{Nomura:2011rb,Nomura:2011dt}:
\begin{itemize}
\item
For a single description allowing a semiclassical interpretation of the 
system, the spacetime region represented is restricted to the causal 
patch associated with a single worldline.  With this restriction, the 
description can be local in the sense that any physical correlations 
between low energy field theoretic degrees of freedom respect causality 
in spacetime (beyond some microscopic quantum gravitational distance 
$l_*$, meaning that possible nonlocal corrections are exponentially 
suppressed $\sim e^{-r/l_*}$).
\end{itemize}
Depending on the worldline we take, we may obtain different descriptions 
of the same system, which are all local in appropriate spacetime regions. 
A transformation between different descriptions is nothing but the 
complementarity transformation.

To implement Hamiltonian quantum mechanics, we must introduce a time 
variable.  This corresponds to foliating the causal patch by equal-time 
hypersurfaces, with a state vector $\ket{\Psi(t)}$ representing the 
state of the system on each hypersurface.%
\footnote{In general, the ``time variable'' of (constrained) Hamiltonian 
 quantum mechanics may not be related directly with time we observe in 
 nature~\cite{DeWitt:1967yk}.  Indeed, the whole ``multiverse'' may be 
 represented by a state that does not depend on the time variable and 
 is normalizable in an appropriate sense~\cite{Nomura:2012zb}.  Even if 
 this is the case, however, when we describe only a branch of the whole 
 state, e.g.\ when we describe a system seen by a particular observer, 
 the state {\it of the system} may depend on time.  In this paper, 
 we discuss systems with black holes, which are {\it parts of} the 
 multiverse so their states may depend on time.}
Let ${\bf x}$ be spatial coordinates parameterizing each equal-time 
hypersurface.  Physical quantities associated with field theoretic 
degrees of freedom can then be obtained using field theoretic operators 
$\phi({\bf x})$ and the state $\ket{\Psi(t)}$.  (Excited string degrees 
of freedom will require the corresponding operators.)  In general, the 
{\it procedure} of electing coordinates $(t,{\bf x})$, which we need to 
{\it define} states and operators, must be given independently of the 
background spacetime, since we do not know it a~priori (and states may 
even represent superpositions of very different semiclassical geometries); 
an example of such procedures is described in Ref.~\cite{Nomura:2013nya}. 
In our discussions in this paper, however, we mostly consider issues 
addressed on a fixed background spacetime (at least approximately), 
so we need not be concerned with this problem too much---we may simply 
use any coordinate system adapted to a particular spacetime we consider, 
e.g.\ Schwarzschild-like coordinates for a black hole.

In the next two sections, we discuss how the complementarity picture 
described above works for a dynamical black hole.  We discuss the 
semiclassical descriptions of the system in various reference frames, 
as well as their mutual consistency.  In these discussions, we focus 
on a black hole that is well approximated by a Schwarzschild black 
hole in asymptotically flat spacetime.  We do not expect difficulty 
in extending it to more general cases.

\section{Black Hole---A Distant Description}
\label{sec:distant}

Suppose we describe the formation and evaporation of a black hole in 
a distant reference frame.  Following Ref.~\cite{'tHooft:1990fr}, we 
postulate that there exists a unitary description which involves only 
the degrees of freedom that can be viewed as being on and outside 
the (stretched) horizon.  To describe quantum states with a black 
hole, we adopt Schwarzschild-like time slicings to define equal-time 
hypersurfaces.%
\footnote{Strictly speaking, to describe a general gravitating system 
 we need a procedure to foliate the relevant spacetime region in a 
 background independent manner, as discussed in the previous section. 
 For our present purposes, however, it suffices to employ any foliation 
 that reduces to Schwarzschild-like time slicings when the black 
 hole exists.  Note that macroscopic uncertainties in the black hole 
 mass, location, and spin caused by the stochastic nature of Hawking 
 radiation~\cite{Page:1979tc,Nomura:2012cx} require us to focus on 
 appropriate branches in the full quantum state in which the black 
 hole in a given time has well-defined values for these quantities 
 at the classical level.  The relation between the Schwarzschild-like 
 foliation and a general background independent foliation is then given 
 by the standard coordinate transformation, which does not introduce 
 subtleties beyond those discussed in this paper.  The effect on unitarity 
 by focusing on particular branches in this way is also minor, so we 
 ignore it.  The full unitarity, however, can be recovered by keeping 
 all the branches in which the black hole has different classical 
 properties at late times~\cite{Nomura:2012cx}. \label{ft:stochastic}}
We argue that the origin of the Bekenstein-Hawking entropy may be 
viewed as a coarse-graining performed to obtain a semiclassical 
description of the evolving black hole.  We then discuss implications 
of such a coarse-graining, in particular how it reconciles unitarity 
of the Hawking emission and black hole mining processes in the 
fundamental theory with the non-unitary (thermal) view in the 
semiclassical description.

\subsection{Microscopic structure of a dynamical black hole}
\label{subsec:micro}

Consider a quantum state which represents a black hole of mass $M$ 
located at some place at rest, where the position and velocity are 
measured with respect to some distant reference frame, e.g.\ an inertial 
frame elected at asymptotic infinity.  Because of the uncertainty 
principle, such a state must involve a superposition of energy and 
momentum eigenstates.  Let us first estimate the required size of the 
spread of energy $\varDelta E$, with $E$ measured in the asymptotic 
region.  According to the standard Hawking calculation, a state of a 
black hole of mass $M$ will evolve after Schwarzschild time $\varDelta t 
\approx O(M l_{\rm P}^2)$ into a state representing a Hawking quantum 
of energy $\approx O(1/M l_{\rm P}^2)$ and a black hole with the 
correspondingly smaller mass.  The fact that these two states---before 
and after the emission---are nearly orthogonal implies that the original 
state must involve a superposition of energy eigenstates with 
\begin{equation}
  \varDelta E \approx \frac{1}{\varDelta t} 
    \approx O\biggl(\frac{1}{M l_{\rm P}^2}\biggr).
\label{eq:delta-E}
\end{equation}
Of course, this is nothing but the standard time-energy uncertainty 
relation, and here we have assumed that a state after time $t \ll 
M l_{\rm P}^2$ is not clearly distinguishable from the original one, 
so that the uncertainty relation is almost saturated.

Next, we consider the spread of momentum $\varDelta p$, where $p$ is 
again measured in the asymptotic region.  Suppose we want to identify 
the spatial location of the black hole with precision comparable to the 
quantum stretching of the horizon $\varDelta r \approx O(1/M)$, i.e.\ 
$\varDelta d \approx O(l_{\rm P})$, where $r$ and $d$ are the Schwarzschild 
radial coordinate and the proper length, respectively.  This implies 
that the superposition must involve momenta with spread $\varDelta p 
\approx (1/M l_{\rm P}) (1/\varDelta d) \approx O(1/M l_{\rm P}^2)$, 
where the factor $1/M l_{\rm P}$ in the middle expression is the redshift 
factor.  This value of $\varDelta p$ corresponds to an uncertainty 
of the kinetic energy $\varDelta E_{\rm kin} \approx p \varDelta p/M 
\approx O(1/M^3 l_{\rm P}^4)$, which is much smaller than $\varDelta E$ 
in Eq.~(\ref{eq:delta-E}).  The spread of energy thus comes mostly 
from a superposition of different rest masses:\ $\varDelta E \approx 
\varDelta M$.

How many different independent ways are there to superpose the energy 
eigenstates to arrive at the same black hole geometry, at a fixed position 
within the precision specified by $\varDelta r$ and of mass $M$ within 
an uncertainty of $\varDelta M$?  We assume that the Bekenstein-Hawking 
entropy, ${\cal A}/4 l_{\rm P}^2$, gives the logarithm of this number (at 
the leading order in expansion in inverse powers of ${\cal A}/l_{\rm P}^2$), 
where ${\cal A} = 16\pi M^2 l_{\rm P}^4$ is the area of the horizon. 
While the definition of the Bekenstein-Hawking entropy does not depend 
on the precise values of $\varDelta M$ or $\varDelta p$, a natural 
choice for these quantities is
\begin{equation}
  \varDelta M \approx \varDelta p 
    \approx O\biggl(\frac{1}{M l_{\rm P}^2}\biggr),
\label{eq:Delta_pE}
\end{equation}
which we will adopt.  The nonzero Bekenstein-Hawking entropy thus implies 
that there are exponentially many independent states in a small energy 
interval of $\varDelta E \approx O(1/M l_{\rm P}^2)$.  We stress that 
it is not appropriate to interpret this to mean that quantum mechanics 
introduces exponentially large degeneracies that do not exist in classical 
black holes.  In classical general relativity, a set of Schwarzschild 
black holes located at some place at rest are parameterized by a continuous 
mass parameter $M$; i.e., there are a continuously infinite number of black 
hole states in the energy interval between $M$ and $M + \varDelta M$ for 
any $M$ and small $\varDelta M$.  Quantum mechanics {\it reduces} this 
to a finite number $\approx e^{S_0} \varDelta M/M$, with $S_0$ given by%
\footnote{Of course, quantum mechanics allows for a superposition of 
 these finite number of independent states, so the number of possible 
 (not necessarily independent) states is continuously infinite.  The 
 statement here applies to the number of independent states, regarding 
 classical black holes with different $M$ as independent states.}
\begin{equation}
  S_0 = \frac{{\cal A}}{4 l_{\rm P}^2} 
    + O\biggl( \frac{{\cal A}^q}{l_{\rm P}^{2q}};\, q < 1 \biggr).
\label{eq:S_0}
\end{equation}
This can also be seen from the fact that $S_0$ is written as ${\cal A} 
c^3/4 l_{\rm P}^2 \hbar$ when $\hbar$ and $c$ are restored, which becomes 
infinite for $\hbar \rightarrow 0$.

As is clear from the argument above, there are exponentially many 
independent microstates, corresponding to Eq.~(\ref{eq:S_0}), which 
are all black hole {\it vacuum} states:\ the states that do not have 
a field or string theoretic excitation on the semiclassical black hole 
background and in which the stretched horizon, located at $r_{\rm s} 
= 2Ml_{\rm P}^2 + O(1/M)$, is not excited.%
\footnote{These states can be defined, for example, as the states obtained 
 by first forming a black hole of mass $M$ and then waiting sufficiently 
 long time after (artificially) switching off Hawking emission.  Note 
 that at the level of full quantum gravity, all the black hole states 
 are obtained as excited states.  Any semiclassical description, however, 
 treats some of them as vacuum states on the black hole background.}
Denoting the indices representing these exponentially many states 
collectively by $k$, which we call the {\it vacuum index}, basis 
states for the general microstates of a black hole of mass $M$ 
(within the uncertainty of $\varDelta M$) can be given by
\begin{equation}
  \ket{\Psi_{\bar{a}\, a\, a_{\rm far}; k}(M)}.
\label{eq:states}
\end{equation}
Here, $\bar{a}$, $a$, and $a_{\rm far}$ represent the indices labeling 
the excitations of the stretched horizon, in the near exterior zone 
region (i.e.\ the region within the gravitational potential barrier 
defined, e.g., as $r \leq R_{\rm Z} \equiv 3M l_{\rm P}^2$), and outside 
the zone ($r > R_{\rm Z}$), respectively.%
\footnote{Strictly speaking, the states may also have the vacuum index 
 associated with the ambient space in which the black hole exists.  The 
 information in this index, however, is not extracted in the Hawking 
 evaporation or black hole mining process, so we ignore it here.  (For 
 more discussions, see, e.g., Section~5 of Ref.~\cite{Nomura:2014yka}.) 
 We will also treat excitations spreading both in the $r \leq R_{\rm Z}$ 
 and $r > R_{\rm Z}$ regions only approximately by including them 
 either in $a$ or $a_{\rm far}$.  The precise description of these 
 excitations will require more elaborate expressions, e.g.\ than the 
 one in Eq.~(\ref{eq:states-2}), which we believe is an inessential 
 technical subtlety in addressing our problem.}
As we have argued, the index $k$ runs over $1, \cdots, e^{S_0}$ for 
the vacuum states $\bar{a} = a = a_{\rm far} =0$.  In general, the range 
for $k$ may depend on $\bar{a}$ and $a$, but its dependence is higher 
order in $l_{\rm P}^2/{\cal A}$; i.e., for fixed $\bar{a}$ and $a$
\begin{equation}
  k = 1,\cdots,e^{S_{\bar{a}a}};
  \qquad S_{\bar{a}a} - S_0 \approx 
    O\biggl( \frac{{\cal A}^q}{l_{\rm P}^{2q}};\, q < 1 \biggr).
\label{eq:k}
\end{equation}
We thus mostly ignore this small dependence of the range of $k$ on 
$(\bar{a},a)$, i.e.\ the non-factorizable nature of the Hilbert space 
factors spanned by these indices, except when we discuss negative energy 
excitations associated with Hawking emission later, where this aspect 
plays a relevant role in addressing one of the firewall arguments.

Since we are mostly interested in physics associated with the black 
hole region, we also introduce the notation in which the excitations 
in the far exterior region are separated.  As we will see later, the 
degrees of freedom represented by $k$ can be regarded as being mostly 
in the region $r \leq R_{\rm Z}$, so we may write the states of the 
entire system in Eq.~(\ref{eq:states}) as
\begin{equation}
  \ket{\Psi_{\bar{a}\, a\, a_{\rm far}; k}(M)} 
  \approx \ket{\psi_{\bar{a} a; k}(M)} \ket{\phi_{a_{\rm far}}(M)},
\label{eq:states-2}
\end{equation}
and call $\ket{\psi_{\bar{a} a; k}(M)}$ and $\ket{\phi_{a_{\rm far}}(M)}$ 
as the black hole and exterior states, respectively.  Note that by labeling 
the states in terms of localized excitations, we need not write explicitly 
the trivial vacuum entanglement between the black hole and exterior states 
that does not depend on $k$, which typically exist when they are specified 
in terms of the occupation numbers of modes spanning the entire space.

How many independent quantum states can the black hole region support? 
Let us label appropriately coarse-grained excitations in the region 
$r_{\rm s} \leq r \leq R_{\rm Z}$ by $i = 1,2,\cdots$, each of which 
carries entropy $S_i$.  Suppose there are $n_i$ excitations of type $i$ 
at some fixed locations.  The entropy of such a configuration is given 
by the sum of the ``entropy of vacuum'' in Eq.~(\ref{eq:S_0}) and the 
entropies associated with the excitations:
\begin{equation}
  S_I = S_0 + \sum_i n_i S_i.
\label{eq:S_I}
\end{equation}
The energy of the system in the region $r \leq R_{\rm Z}$ is given 
by the sum of the mass $M$ of the black hole, which we define as the 
energy the system would have in the absence of an excitation outside 
the stretched horizon, and the energies associated with the excitations 
in the zone.  Note that excitations here are defined as fluctuations 
with respect to a fixed background, so their energies $E_i$ as well 
as entropies $S_i$ can be either positive or negative, although the 
signs of the energy and entropy must be the same:\ $E_i S_i > 0$. 
The meaning of negative entropies will be discussed in detail in 
Sections~\ref{subsec:Hawking} and \ref{subsec:mining}.

Since excitations in the zone affect geometry, spacetime outside 
the stretched horizon, when they exist, is not exactly that of a 
Schwarzschild black hole.  We require that these excitations do not 
form a black hole by themselves or become a part of the black hole at 
the center; otherwise, the state must be viewed as being built on a 
different semiclassical vacuum.%
\footnote{More precisely, we regard two geometries as being built on 
 different classes of semiclassical vacua when they have different horizon 
 configurations as viewed from a fixed reference frame.  On the other 
 hand, if two geometries have the same horizon, they belong to the same 
 ``vacuum equivalence class'' in the sense that one can be converted into 
 the other with ``excitations.''  For more discussions on this point, 
 see Ref.~\cite{Nomura:2013nya} and Section~\ref{subsec:semiclassical}.}
The total entropy $S$ of the region $r \leq R_{\rm Z}$, i.e.\ the number 
of independent microscopic quantum states representing this region, 
is then given by
\begin{equation}
  S = \ln \Bigl( \sum_I e^{S_I} \Bigr),
\label{eq:S-def}
\end{equation}
where $I$ represents possible configurations of excitations, specified by 
the set of numbers $\{ n_i \}$ and the locations of excitations of each 
type $i$, that do not modify the semiclassical vacuum in the sense described 
above.  As suggested by a representative estimate~\cite{'tHooft:1993gx}, 
and particularly emphasized in Ref.~\cite{Nomura:2013lia}, the contribution 
of such excitations to the total entropy is subdominant in the expansion 
in inverse powers of ${\cal A}/l_{\rm P}^2$: $S = S_0 + O( {\cal A}^q 
/ l_{\rm P}^{2q}; q < 1)$.  The total entropy in the near black hole 
region, $r \leq R_{\rm Z}$, is thus given by
\begin{equation}
  S = \frac{{\cal A}}{4 l_{\rm P}^2},
\label{eq:S}
\end{equation}
at the leading order in $l_{\rm P}^2/{\cal A}$.

\subsection{Emergence of the semiclassical picture and coarse-graining}
\label{subsec:semiclassical}

The fact that all the independent microstates with different values of 
$k$ lead to the same geometry suggests that the semiclassical picture 
is obtained after coarse-graining the degrees of freedom represented 
by this index; namely, any result in semiclassical theory is a statement 
about the maximally mixed ensemble of microscopic quantum states 
consistent with the specified background within the precision allowed 
by quantum mechanics~\cite{Nomura:2014yka}.  According to this picture, 
the black hole vacuum state in the semiclassical description is given 
by the density matrix
\begin{equation}
  \rho_0(M) = \frac{1}{e^{S_0}} \sum_{k=1}^{e^{S_0}} 
    \ket{\Psi_{\bar{a}=a=a_{\rm far}=0; k}(M)} 
    \bra{\Psi_{\bar{a}=a=a_{\rm far}=0; k}(M)}.
\label{eq:rho_0}
\end{equation}
Because of the coarse-graining of an enormous number of degrees of 
freedom, this density matrix has statistical characteristics.

In order to obtain the response of this state to the operators in the 
semiclassical theory, we may trace out the subsystem on which they 
do not act.  As we will discuss more later, the operators in the 
semiclassical theory in general act on a part, but not all, of the 
degrees of freedom represented by the $k$ index.  Let us denote the 
subsystem on which semiclassical operators act nontrivially by $C$, 
and its complement by $\bar{C}$.  The index $k$ may then be viewed 
as labeling the states in the combined $C\bar{C}$ system which 
satisfy certain constraints, e.g.\ the total energy being $M$ within 
$\varDelta M$.  The density matrix representing the semiclassical 
vacuum state in the Hilbert space in which the semiclassical operators 
act nontrivially, $C$, is given by
\begin{equation}
  \tilde{\rho}_0(M) = {\rm Tr}_{\bar{C}}\, \rho_0(M).
\label{eq:rho_0-sc}
\end{equation}
Consistently with our identification of the origin of the Bekenstein-Hawking 
entropy, we assume that this density matrix represents the thermal density 
matrix with temperature $T_{\rm H} = 1/8\pi M l_{\rm P}^2$ in the zone 
region (as measured at asymptotic infinity):
\begin{equation}
  \tilde{\rho}_0(M) \approx \frac{1}{{\rm Tr}\, e^{-\beta H_{\rm sc}(M)}}
    e^{-\beta H_{\rm sc}(M)};
\qquad
  \beta = \left\{ \begin{array}{ll}
    \frac{1}{T_{\rm H}} & \mbox{for } r \leq R_{\rm Z}, \\
    +\infty & \mbox{for } r > R_{\rm Z},
    \end{array} \right.
\label{eq:rho_0-therm}
\end{equation}
where $H_{\rm sc}(M)$ is the Hamiltonian of the semiclassical theory 
in the distant reference frame, which is defined in the region $r \geq 
r_{\rm s}$ on the black hole background of mass $M$.%
\footnote{The Hilbert space of the semiclassical theory for states which 
 have a single black hole at a fixed location at rest may be decomposed 
 as ${\cal H} = \oplus_M {\cal H}_M$, where ${\cal H}_M$ is the space 
 spanned by the states in which there is a black hole of (appropriately 
 coarse-grained) mass $M$.  In this language, $H_{\rm sc}(M)$ is a part 
 of the semiclassical Hamiltonian acting on the subspace ${\cal H}_M$.}
(The meaning of position-dependent $\beta$ is that the expression 
$\beta H_{\rm sc}(M)$ should be interpreted as $\beta$ times the 
Hamiltonian density integrated over space.)  Note that this procedure 
of obtaining Eq.~(\ref{eq:rho_0-therm}) from Eq.~(\ref{eq:rho_0}) 
can be viewed as an example of the standard procedure of obtaining 
the canonical ensemble of a system from the microcanonical ensemble 
of a larger (isolated) system that contains the system of interest. 
In fact, if the system traced out is larger than the system of 
interest, ${\rm dim}\,\bar{C} \gtrsim {\rm dim}\,C$, we expect to 
obtain the canonical ensemble in this manner (see Ref.~\cite{Page:1993df} 
for a related discussion).  Below, we drop the tilde from the density 
matrix in Eq.~(\ref{eq:rho_0-therm}), as it represents the same state 
as the one in Eq.~(\ref{eq:rho_0})---$\rho_0(M)$ must be interpreted 
to mean either the right-hand side of Eq.~(\ref{eq:rho_0}) or of 
Eq.~(\ref{eq:rho_0-therm}), depending on the Hilbert space under 
consideration.

In semiclassical field theory, the density matrix of 
Eq.~(\ref{eq:rho_0-therm}) is obtained as a reduced density 
matrix by tracing out the region within the horizon in the {\it unique} 
global black hole vacuum state.  Our view is that this density matrix, 
in fact, is obtained from a mixed state of exponentially many pure 
states, arising from a coarse-graining performed in Eq.~(\ref{eq:rho_0}); 
the prescription in the semiclassical theory provides (merely) a useful 
way of obtaining the same density matrix, in a similar sense in which 
the thermofield double state was originally introduced~\cite{TFD}. 
We emphasize that the information in $k$ is invisible in the semiclassical 
theory (despite the fact that it involves subsystem $C$) as it is 
already coarse-grained {\it to obtain} the theory; in particular, 
the dynamics of the degrees of freedom represented by $k$ cannot be 
described in terms of the semiclassical Hamiltonian $H_{\rm sc}(M)$.%
\footnote{This does not mean that a device made out of semiclassical 
 degrees of freedom cannot probe information in $k$.  Since there 
 are processes in the fundamental theory (i.e.\ Hawking evaporation 
 and mining processes) in which information in $k$ is transferred to 
 that in semiclassical {\it excitations} (i.e.\ degrees of freedom 
 represented by the $a$ and $a_{\rm far}$ indices), information in 
 $k$ can be probed by degrees of freedom appearing in the semiclassical 
 theory.  It is simply that these information extraction processes 
 cannot be described within the semiclassical theory, since it can 
 make statements only about the ensemble in Eq.~(\ref{eq:rho_0}) 
 and excitations built on it.}
As we will see explicitly later, it is this inaccessibility of $k$ 
that leads to the apparent violation of unitarity in the semiclassical 
calculation of the Hawking emission process~\cite{Hawking:1976ra}. 
Note that because $\rho_0(M)$ takes the form of the maximally mixed 
state in $k$, results in the semiclassical theory do not depend on 
the basis of the microscopic states chosen in this space.

A comment is in order.  In connecting the expression in Eq.~(\ref{eq:rho_0}) 
to Eq.~(\ref{eq:rho_0-therm}), we have (implicitly) assumed that 
$\ket{\Psi_{\bar{a}=a=a_{\rm far}=0; k}(M)}$ represent the black 
hole vacuum states in the limit that the effect from evaporation 
is (artificially) shut off.%
\footnote{This is analogous to the treatment of a meta-stable vacuum 
 in usual quantum field theory.  At the most fundamental level (or on 
 a very long timescale), such a state must be viewed as a scattering 
 state built on the true ground state of the system.  In practice (or 
 on a sufficiently short timescale), however, we regard it as a vacuum 
 state, which is approximately the ground state of a theory in which 
 the tunneling out of this state is artificially switched off, e.g.\ 
 by making the relevant potential barriers infinitely high.}
With this definition of vacuum states, the evolution effect necessarily 
``excites'' the states, making $a \neq 0$, as we will see more explicitly 
in Section~\ref{subsec:Hawking}.  As a consequence, the density matrix 
for the semiclassical operators representing the evolving black hole 
deviates from Eq.~(\ref{eq:rho_0-therm}) even without matter or radiation. 
(In the semiclassical picture, this is due to the fact that the effective 
gravitational potential is not truly confining, so that the state of 
the black hole is not completely stationary.)  If one wants, one can 
redefine vacuum states to be these states:\ the states that do not 
have any matter or radiation excitation on the {\it evolving} black 
hole background---the original vacuum states are then obtained as 
excited states on the new vacuum states.%
\footnote{In the standard language in semiclassical theory, the 
 original vacuum states correspond essentially to the Hartle-Hawking 
 vacuum~\cite{Hartle:1976tp}, while the new ones (very roughly) to 
 the Unruh vacuum~\cite{Unruh:1976db}.}
This redefinition is possible because the two semiclassical ``vacua'' 
represented by the two classes of microstates belong to the same 
``vacuum equivalence class'' in the sense described in the last 
paragraph of Section~\ref{subsec:micro}; specifically, they possess 
the same horizon for the same black hole mass, as defined for the 
evaporating case in Ref.~\cite{Bardeen:1981zz}.

As was mentioned above, semiclassical operators, in particular those 
for modes in the zone, act nontrivially on both $a$ and $k$ indices 
of microstates $\ket{\Psi_{\bar{a}\, a\, a_{\rm far}; k}(M)}$.  This 
can be seen as follows.  If the operators acted only on the $a$ index, 
the maximal mixture in $k$ space with $a = 0$, Eq.~(\ref{eq:rho_0}), 
would look like a pure state from the point of view of these operators, 
contradicting the thermal nature in Eq.~(\ref{eq:rho_0-therm}).  On 
the other hand, if the operators acted only on the $k$ index, they would 
commute with the maximally mixed state in $k$ space, again contradicting 
the thermal state.  Since the thermal nature of Eq.~(\ref{eq:rho_0-therm}) 
is prominent only for modes whose energies as measured in the asymptotic 
region are of order the Hawking temperature or smaller
\begin{equation}
  \omega \lesssim T_{\rm H},
\label{eq:IR-modes}
\end{equation}
i.e.\ whose energies as measured by local (approximately) static 
observers are of order or smaller than the blueshifted Hawking temperature 
$T_{\rm H}/\sqrt{1-2Ml_{\rm P}^2/r}$, this feature is significant only 
for such infrared modes---operators representing modes with $\omega 
\gg T_{\rm H}$ act essentially only on the $a$ index.  For operators 
representing the modes with Eq.~(\ref{eq:IR-modes}), their actions 
on microstates can be very complicated, although they act on the 
coarse-grained vacuum state of Eq.~(\ref{eq:rho_0}) as if it is the 
thermal state in Eq.~(\ref{eq:rho_0-therm}), up to corrections suppressed 
by the exponential of the vacuum entropy $S_0$.  The commutation relations 
of these operators defined on the coarse-grained states take the form 
as in the semiclassical theory, again up to exponentially suppressed 
corrections.

There is a simple physical picture for this phenomenon of 
``non-decoupling'' of the $a$ and $k$ indices for the infrared modes. 
As viewed from a distant reference frame, these modes are ``too soft'' 
to be resolved clearly above the background---since the derivation 
of the semiclassical theory involves coarse-graining over microstates 
in which the energy stored in the region $r \lesssim R_{\rm Z}$ 
has spreads of order $\varDelta E \approx 1/M l_{\rm P}^2$, infrared 
modes with $\omega \lesssim T_{\rm H} \approx O(1/M l_{\rm P}^2)$ are 
not necessarily distinguished from ``spacetime fluctuations'' 
of order $\varDelta E$.  One might think that if a mode has nonzero 
angular momentum or charge, one can discriminate it from spacetime 
fluctuations.  In this case, however, it cannot be clearly distinguished 
from vacuum fluctuations of a Kerr or Reissner-Nordstr\"{o}m black 
hole having the corresponding (minuscule) angular momentum or charge. 
In fact, we may reverse the logic and view that this lack of a clear 
identity of the soft modes is the physical origin of the thermality 
of black holes (and thus of Hawking radiation).

Once the state for the vacuum of the semiclassical theory is obtained 
as in Eq.~(\ref{eq:rho_0}) (or Eq.~(\ref{eq:rho_0-therm}) after partial 
tracing) and appropriate coarse-grained operators acting on it are 
identified, it is straightforward to construct the rest 
of the states in the theory---we simply have to act these operators (either 
field theoretic or of excited string states) on $\rho_0(M)$ to obtain the 
excited states.  For example, to obtain a state which has a field theoretic 
excitation in the zone, one can apply the appropriate linear combination 
of creation and/or annihilation operators in the semiclassical theory, 
$a^\dagger_{\omega \ell m}$ and/or $a_{\omega \ell m}$:
\begin{equation}
  \rho_{\bar{a}=0\, a\, a_{\rm far}=0}(M) 
  = \left( \sum_{\ell,m} \int (c^a_{\omega \ell m} 
      a_{\omega \ell m} + c^{\prime a}_{\omega \ell m} 
      a^\dagger_{\omega \ell m}) d\omega \right) \rho_0(M) 
    \left( \sum_{\ell,m} \int (c^a_{\omega \ell m} 
      a_{\omega \ell m} + c^{\prime a}_{\omega \ell m} 
      a^\dagger_{\omega \ell m}) d\omega \right)^\dagger,
\label{eq:rho_a}
\end{equation}
where $c^a_{\omega \ell m}$ and $c^{\prime a}_{\omega \ell m}$ are 
coefficients.  In the case that the applied operator is that for an 
infrared mode, this represents a state in which the thermal distribution 
for the infrared modes is ``modulated'' by an excitation over it. 
A construction similar to Eq.~(\ref{eq:rho_a}) also works for excitations 
in the far region.  To obtain excitations of the stretched horizon, 
i.e.\ $\bar{a} \neq 0$, operators dedicated to describing them must 
be introduced.  The detailed dynamics of these degrees of freedom, 
i.e.\ the $r = r_{\rm s}$ part of $H_{\rm sc}(M)$, is not yet fully 
known, however.

\subsection{``Constituents of spacetime'' and their distribution}
\label{subsec:const}

While not visible in semiclassical theory, the black hole formation 
and evaporation (or mining) processes do involve the degrees of freedom 
represented by $k$, which we call {\it fine-grained vacuum degrees of 
freedom}, or vacuum degrees of freedom for short.  The dynamics of these 
degrees of freedom as well as their interactions with the excitations 
in the semiclassical theory are determined by the fundamental theory 
of quantum gravity, which is not yet well known.  We may, however, 
anticipate their basic properties based on some general considerations. 
In particular, motivated by the general idea of complementarity, we 
assume the following:
\begin{itemize}
\item
Interactions with vacuum degrees of freedom do not introduce violation 
of causality among field theory degrees of freedom (except possibly 
for exponentially suppressed corrections, $\sim e^{-r/l_*}$ with $l_*$ 
a short-distance quantum gravitational scale).
\item
Interactions between vacuum degrees of freedom and excitations in 
the semiclassical theory are such that unitarity is preserved at the 
microscopic level.
\end{itemize}
The first assumption is a special case of the postulate discussed 
in Section~\ref{sec:failure}, applied to the distant reference frame 
description of a black hole.  This implies that we cannot send 
superluminal signals among field theory degrees of freedom using 
interactions with vacuum degrees of freedom.  The second assumption 
has an implication for how the vacuum degrees of freedom may appear 
from the semiclassical standpoint, which we now discuss.

In quantum mechanics, the information about a state is generally 
delocalized in space---locality is a property of dynamics, not that 
of states.  In the case of black hole states, the information about $k$, 
which roughly represents slightly different ``values'' (superpositions) 
of $M$, is generally delocalized in a large spatial region, so that it 
can be accessed physically in a region away from the stretched horizon 
(e.g.\ around the edge of the zone $r \sim R_{\rm Z}$).  This, however, 
does not mean that the complete information about the state can be 
recovered by a physical process occurring in a limited region in spacetime. 
For example, if we consider the set of $e^{S_0}$ different black hole 
vacuum states, a physical detector occupying a finite spatial region 
can only partially discriminate these states in a given finite time.

To see how much information a physical detector in spatial region $i$ 
can resolve, we can consider the reduced density matrix obtained after 
tracing out the subsystems that cannot be accessed by the semiclassical 
degrees of freedom associated with this region.  In particular, we may 
consider the set of all field theory (and excited string state) operators 
that have support in $i$, and trace out the subsystems that do not 
respond to any of these operators (which we denote by $\bar{C}_i$):
\begin{equation}
  \rho_0^{(i)} = {\rm Tr}_{\bar{C}_i}\, \rho_0(M),
\label{eq:rho_0-i}
\end{equation}
where $\rho_0(M)$ is given by Eq.~(\ref{eq:rho_0}), and we have omitted 
the argument $M$ for $\rho_0^{(i)}$.  The von~Neumann entropy of this 
density matrix, $S_0^{(i)} = -{\rm Tr}\, \rho_0^{(i)} \ln \rho_0^{(i)}$, 
then indicates the discriminatory power the region $i$ possesses---a 
physical process occurring in region $i$ can, at most, discriminate 
the $e^{S_0}$ states into $e^{S_0^{(i)}}$ ($\ll e^{S_0}$) types 
in a characteristic timescale of the system, $1/\varDelta E \approx 
O(M l_{\rm P}^2)$.  According to the assumption in Eq.~(\ref{eq:rho_0-therm}), 
this entropy is the gravitational thermal entropy contained in region 
$i$, calculated using the semiclassical theory.

We therefore arrive at the following picture.  Let us divide the region 
$r \geq r_{\rm s}$ into $N$ (arbitrary) subregions, each of which is 
assumed to have a sufficiently large number of degrees of freedom so 
that the thermodynamic limit can be applied.  A basis state in the 
semiclassical theory can be written as
\begin{equation}
  \rho_{\bar{a}\, a\, a_{\rm far}}(M) 
  = \rho^{(1)}_{a_1} \otimes \rho^{(2)}_{a_2} \otimes 
    \cdots \otimes \rho^{(N)}_{a_N},
\label{eq:rho-decomp}
\end{equation}
where $\rho^{(i)}_{a_i}$ are states defined in the $i$-th subregion, with 
$a_i$ representing excitations contained in that region.  (Following the 
convention in Section~\ref{subsec:semiclassical}, we regard the vacuum 
states, $\bar{a} = a = a_{\rm far} = 0$, to be defined in the limit that 
the effect from evaporation is ignored.)  Now, in the full Hilbert space 
of quantum gravity, there are $e^{S_0}$ independent states that all reduce 
to the same $\rho_{\bar{a}\, a\, a_{\rm far}}(M)$ at the semiclassical 
level.  These states can be written as
\begin{equation}
  \ket{\Psi_{\bar{a}\, a\, a_{\rm far}; k = \{ k_i \}}(M)} 
  = \ket{\psi^{(1)}_{a_1; k_1}}\, \ket{\psi^{(2)}_{a_2; k_2}} 
    \cdots \ket{\psi^{(N)}_{a_N; k_N}},
\label{eq:Psi_k-decomp}
\end{equation}
where $k_i = 1,\cdots,e^{S_0^{(i)}}$ with
\begin{equation}
  S_0^{(i)} \,\approx\, 
    \mbox{gravitational thermal entropy contained in subregion } i,
\label{eq:distr}
\end{equation}
calculated using the semiclassical theory for subregions that do not 
contain the stretched horizon.  The $S_0^{(i)}$'s for the subregions 
involving the stretched horizon are determined by the condition
\begin{equation}
  \sum_{i=1}^{N} S_0^{(i)} = S_0 \approx \frac{{\cal A}}{4 l_{\rm P}^2},
\label{eq:S_0-i-sum}
\end{equation}
which is valid in the thermodynamic limit.  Assuming that the entropy 
on the stretched horizon is distributed uniformly on the surface, this 
condition determines the entropies contained in all the subregions.

The association of $k_i$'s to each subregion, as in 
Eq.~(\ref{eq:Psi_k-decomp}), corresponds to taking a specific basis 
in the space spanned by $k$.  While the expressions above are strictly 
valid only in the thermodynamic limit, the corrections caused by deviating 
from it (e.g.\ due to correlations among subregions) do not affect 
our later discussions.  In particular, it does not change the fact 
that the region around the edge of the zone, $r \leq R_{\rm Z}$ and 
$r - 2Ml_{\rm P}^2 \,\slashed{\ll}\, Ml_{\rm P}^2$, contains $O(1)$ 
bits of information about $k$ (as it contains $O(1)$ bits of gravitational 
thermal entropy), which becomes important when we discuss the Hawking 
emission process in Section~\ref{subsec:Hawking}.  Incidentally, the 
picture described here leads to the natural interpretation that the 
subsystem that is traced out when going from Eq.~(\ref{eq:rho_0}) to 
Eq.~(\ref{eq:rho_0-therm}) corresponds to the stretched horizon; i.e.\ 
$\bar{C}$ lives on the stretched horizon, while $C$ in the zone.%
\footnote{This in turn gives us a natural prescription to determine 
 the location of the stretched horizon precisely.  Since the semiclassical 
 expression in Eq.~(\ref{eq:rho_0-therm}) is expected to break down for 
 $\ln {\rm dim}\,C > \ln {\rm dim}\,\bar{C}$, a natural place to locate 
 the stretched horizon, i.e.\ the cutoff of the semiclassical spacetime, 
 is where the gravitational thermal entropy outside the stretched horizon 
 becomes $S_0/2 = {\cal A}/8 l_{\rm P}^2$.  For $n$ low energy species, 
 this yields $r_{\rm s} - 2Ml_{\rm P}^2 \sim n/M \sim l_*^2/M l_{\rm P}^2$, 
 where $l_*$ is the string (cutoff) scale and we have used the relation 
 $l_*^2 \sim n l_{\rm P}^2$, which is expected to apply in any consistent 
 theory of quantum gravity (see, e.g., Ref.~\cite{Dvali:2007hz}).  This 
 scaling is indeed consistent, giving the local Hawking temperature at 
 the stretched horizon $T(r_{\rm s}) \sim 1/l_*$, where $T(r)$ is given 
 in Eq.~(\ref{eq:T_local}). \label{ft:stretched}}

We stress that by the gravitational thermal entropy in Eq.~(\ref{eq:distr}), 
we mean that associated with the equilibrium vacuum state.  It counts 
the thermal entropy within the zone, since this region is regarded as 
being in equilibrium because of its boundedness due to the stretched 
horizon and the potential barrier; on the other hand, Eq.~(\ref{eq:distr}) 
does not count the thermal entropy associated with Hawking radiation 
emitted from the zone, which is (artificially) switched off in defining 
our vacuum microstates.  In other words, when calculating $S_0^{(i)}$'s 
using Eq.~(\ref{eq:distr}) we should use the vacuum state in 
Eq.~(\ref{eq:rho_0-therm}), implying that we should use the local 
temperature, i.e.\ the temperature as measured by local static 
observers, of
\begin{equation}
  T(r) \simeq \left\{ \begin{array}{ll} 
    \frac{T_{\rm H}}{\sqrt{1-\frac{2Ml_{\rm P}^2}{r}}} 
      & \mbox{for } r \leq R_{\rm Z}, \\
    0 & \mbox{for } r > R_{\rm Z}.
    \end{array} \right.
\label{eq:T_local}
\end{equation}
When the evolution effect is turned on, which we will analyze in 
Section~\ref{subsec:Hawking}, the state of the zone is modified 
($a \neq 0$) due to an ingoing negative energy flux, while the state 
outside the zone is excited ($a_{\rm far} \neq 0$) by Hawking quanta, 
which are emitted from the edge of the zone and propagate freely 
in the ambient space.  The contribution of the negative energy 
flux to the entropy within the zone is small, as we will see in 
Section~\ref{subsec:Hawking}.

The distribution of vacuum degrees of freedom in 
Eqs.~(\ref{eq:Psi_k-decomp},~\ref{eq:distr}) is exactly the one 
needed for the interactions between these degrees of freedom and 
semiclassical excitations to preserve unitarity~\cite{Nomura:2014yka}. 
Imagine we put a physical detector at constant $r$ in the zone.  The 
detector then sees the thermal bath for all the modes with blueshifted 
Hawking temperature, Eq.~(\ref{eq:T_local}), including higher angular 
momentum modes.  This allows for the detector(s) to extract energy 
from the black hole at an accelerated rate compared with spontaneous 
Hawking emission:\ the mining process~\cite{Unruh:1982ic,Brown:2012un}. 
In order for this process to preserve unitarity, the detector must also 
extract information at the correspondingly accelerated rate.  This is 
possible if the information about the microstate of the black hole, 
specified by the index $k$, is distributed according to the gravitational 
thermal entropy, as in Eqs.~(\ref{eq:Psi_k-decomp},~\ref{eq:distr}). 
A similar argument also applies to the spontaneous Hawking emission 
process, which is viewed as occurring around the edge of the zone, 
$r \sim R_{\rm Z}$, where the gravitational thermal entropy is 
small but not negligible.  The microscopic and semiclassical 
descriptions of these processes will be discussed in detail in 
Sections~\ref{subsec:Hawking} and \ref{subsec:mining}.

It is natural to interpret the expression in Eq.~(\ref{eq:Psi_k-decomp}) 
to mean that $k_i$ labels possible configurations of ``physical soft 
quanta''---or the ``constituents of spacetime''---that comprise the 
region $i$.  In a certain sense, this interpretation is correct.  The 
dimension of the relevant Hilbert space, $e^{S_0^{(i)}}$, controls 
possible interactions of the vacuum degrees of freedom with the excitations 
in the semiclassical theory in region $i$, e.g.\ how much information 
a detector located in region $i$ can extract from the vacuum degrees 
of freedom.  This simple picture, however, breaks down when we describe 
the same system from a different reference frame.  As we will discuss in 
Section~\ref{sec:infalling}, the distribution of the vacuum degrees of 
freedom {\it depends on the reference frame}---they are not ``anchored'' 
to spacetime.  Nevertheless, in a fixed reference frame, the concept of 
the spatial distribution of the degrees of freedom represented by the 
index $k$ does make sense.  In particular, in a distant reference frame 
the distribution is given by the gravitational thermal entropy calculated 
in the semiclassical theory, as we discussed here.

\subsection{Hawking emission---``microscopic'' and semiclassical 
 descriptions}
\label{subsec:Hawking}

The formation and evaporation of a black hole involve processes in which 
the information about the initial collapsing matter is transferred into 
the vacuum index $k$, which will later be transferred back to the excitations 
in the semiclassical theory, i.e.\ the state of final Hawking radiation. 
Schematically, we may write these processes as
\begin{equation}
  \ket{m_{\rm init}} \,\,\rightarrow\,\, 
    \sum_{k=1}^{e^{S_0(M(t))}} \sum_l\! c_{kl}(t)\, 
      \ket{\psi_k(M(t))}\, \ket{r_l(t)} 
  \,\,\rightarrow\,\, \ket{r_{\rm fin}},
\label{eq:BH-evol}
\end{equation}
where $\ket{m_{\rm init}}$, $\ket{\psi_k(M(t))}$, $\ket{r_l(t)}$, and 
$\ket{r_{\rm fin}}$ represent the states for the initial collapsing 
matter, the black hole of mass $M(t)$ (which includes the near exterior 
zone region; see Eq.~(\ref{eq:states-2})), the subsystem complement to 
the black hole at time $t$, and the final Hawking quanta after the black 
hole is completely evaporated, respectively.  Here, we have suppressed the 
indices representing excitations for the black hole states.  For generic 
initial states and microscopic emission dynamics, this evolution satisfies 
the behavior outlined in Ref.~\cite{Page:1993wv} on general grounds.

In this subsection, we discuss how the black hole evaporating process in 
Eq.~(\ref{eq:BH-evol}) proceeds in details, elucidating how the arguments 
for firewalls in Refs.~\cite{Almheiri:2012rt,Almheiri:2013hfa,Marolf:2013dba} 
are avoided.  We also discuss how the semiclassical theory describes 
the same process, elucidating how the thermality of Hawking radiation 
arises despite the unitarity of the process at the fundamental level.

\subsubsection{``Microscopic'' (unitary) description}

Let us first consider how the ``elementary'' Hawking emission process 
is described at the microscopic level,%
\footnote{By the ``microscopic'' description, we mean a description in 
 which the vacuum index $k$ is kept (i.e.\ not coarse-grained as in the 
 semiclassical description) so that the process is manifestly unitary 
 at each stage of the evolution.  A complete description of the microscopic 
 dynamics of the vacuum degrees of freedom requires the fundamental 
 theory of quantum gravity, which is beyond the scope of this paper.}
i.e.\ how a ``single'' Hawking emission occurs in the absence of any 
excitations other than those directly associated with the emission. 
(As we will see later, this is not a very good approximation in general, 
but the treatment here is sufficient to illustrate the basic mechanism 
by which the information is transferred from the black hole to the 
ambient space.)

Suppose a black hole of mass $M$ is in microstate $k$:
\begin{equation}
  \ket{\Psi_k(M)} = \ket{\psi_k(M)} \ket{\phi_I},
\label{eq:H-before}
\end{equation}
where $\ket{\psi_k(M)}$ is the black hole state, in which we have 
omitted indices representing excitations, while $\ket{\phi_I}$ is the 
exterior state, from which we have suppressed small $M$ dependence (which, 
e.g., causes a small gravitational redshift of a factor of about $1.5$ 
for the emitted Hawking quanta to reach the asymptotic region).  As 
discussed in Sections~\ref{subsec:semiclassical} and \ref{subsec:const}, 
we consider $\ket{\Psi_k(M)}$ to be one of the black hole vacuum 
microstates in the limit that the evolution effect is shut off; see, 
e.g., Eqs.~(\ref{eq:rho_0-therm},~\ref{eq:T_local}).  The effect 
of the evolution, which consists of successive elementary Hawking 
emission processes, will be discussed later.

After a timescale of $t \approx O(M l_{\rm P}^2)$, the state in 
Eq.~(\ref{eq:H-before}) evolves due to Hawking emission as
\begin{equation}
  \ket{\psi_k(M)} \ket{\phi_I} \rightarrow 
    \sum_{i,a,k'} c^k_{i a k'} \ket{\psi_{a; k'}(M)} \ket{\phi_{I+i}},
\label{eq:H-emission}
\end{equation}
where $\ket{\phi_{I+i}}$ is the state in which newly emitted Hawking 
quanta, labeled by $i$ and having total energy $E_i$, are added to the 
appropriately time evolved $\ket{\phi_I}$.  The index $a$ represents 
the fact that the black hole state has negative energy excitations of 
total energy $-E_a$ ($E_a > 0$) around the edge of the zone, created in 
connection with the emitted Hawking quanta; the coefficients $c^k_{i a k'}$ 
are nonzero only if $E_i \approx E_a$ (within the uncertainty).%
\footnote{To be precise, the sum in the right-hand side of 
 Eq.~(\ref{eq:H-emission}) contains the ``$i=0$ terms'' representing 
 the branches in which no quantum is emitted:\ $\ket{\phi_{I+0}} 
 = \ket{\phi_I}$.  In these terms, there is no negative energy 
 excitation:\ $c^k_{0 a k'} \neq 0$ only for $a = 0$.  The following 
 expressions are valid including these terms with the definition 
 $E_{i=0} = E_{a=0} = 0$.}
The negative energy excitations then propagate inward, and after a time 
of order $M l_{\rm P}^2 \ln(M l_{\rm P})$ collide with the stretched 
horizon, making the black hole states relax as
\begin{equation}
  \ket{\psi_{a; k'}(M)} \rightarrow 
    \sum_{k_a} d^{a k'}_{k_a} \ket{\psi_{k_a}(M-E_a)}.
\label{eq:H-relax}
\end{equation}
The combination of Eqs.~(\ref{eq:H-emission},~\ref{eq:H-relax}) yields
\begin{equation}
  \ket{\psi_k(M)} \ket{\phi_I} \rightarrow 
    \sum_{i,k_i} \alpha^k_{i k_i} \ket{\psi_{k_i}(M-E_i)} \ket{\phi_{I+i}},
\label{eq:H-process}
\end{equation}
where $\alpha^k_{i k_i} = \sum_{a,k'} c^k_{i a k'} d^{a k'}_{k_i}$, and 
we have used $E_i = E_a$; here, $M - E_i$ for different $i$ may belong 
to the same mass within the precision $\varDelta M$, i.e.\ $M - E_i 
= M - E_{i'}$ for $i \neq i'$.  This expression shows that information 
in the black hole can be transferred to the radiation state $i$.

It is important that the negative energy excitations generated in 
Eq.~(\ref{eq:H-emission}) come with {\it negative entropies}, so that 
each of the processes in Eqs.~(\ref{eq:H-emission},~\ref{eq:H-relax}) 
(as well as the propagation of the negative energy excitations in 
the zone) is separately unitary.  This means that as $k$ and $i$ run 
over all the possible values with $a$ being fixed, the index $k'$ runs 
only over $1,\cdots,e^{S_0(M-E_a)}$, the dimension of the space spanned 
by $k_a$.  In fact, this is an example of the non-factorizable nature 
of the Hilbert space factors spanned by $k$ and $a$ discussed in 
Eq.~(\ref{eq:k}), which we assume to arise from the fundamental theory. 
This structure of the Hilbert space allows for avoiding the argument 
for firewalls in Ref.~\cite{Almheiri:2013hfa}---unlike what is imagined 
there, elements of the naive Fock space built on each $k$ in a way 
isomorphic to that of quantum field theory are not all physical; the 
physical Hilbert space is smaller than such a (hypothetical) Fock 
space.  This implies, in particular, that the Fock space structure 
of a semiclassical theory does not factor from the space spanned 
by the vacuum index $k$, as is also implied by the analysis in 
Section~\ref{subsec:semiclassical}.

To further elucidate the point made above, we can consider the following 
simplified version of the relevant processes.  Suppose a black hole in 
a superposition state of $\ket{\psi_k(M)}$'s ($k = 1,\cdots,e^{S_0(M)}$) 
releases 1~bit of information through Hawking emission of the form:
\begin{equation}
  \ket{\psi_k(M)} \ket{\phi_0} 
  \rightarrow \left\{ \begin{array}{ll}
    \ket{\psi_{a;\frac{k+1}{2}}(M)} \ket{\phi_1} & 
      \mbox{if $k$ is odd}, \\
    \ket{\psi_{a;\frac{k}{2}}(M)} \ket{\phi_2} &
      \mbox{if $k$ is even},
  \end{array} \right.
\label{eq:H-toy-1}
\end{equation}
where we have assumed $E_1 = E_2 = (\ln 2)/8\pi M l_{\rm P}^2 \simeq 
T_{\rm H}$, so that the entropy of the black hole after the emission 
is reduced by 1~bit:\ $S_0(M-E_1) = S_0(M) - \ln 2$.  Note that the 
index representing the negative energy excitation (of energy $-E_1$) 
takes the same value $a$ in the first and second lines.  Namely, while 
the entire process in Eq.~(\ref{eq:H-toy-1}) is unitary, the initial 
states with $k = 2n-1$ and $2n$ lead to the {\it same black hole state}. 
After the negative energy excitation reaches the stretched horizon, the 
black hole states relax into vacuum states for a smaller black hole:
\begin{equation}
  \ket{\psi_{a;k'}(M)} \rightarrow \ket{\psi_{k_1=k'}(M-E_1)}.
\label{eq:H-toy-2}
\end{equation}
While the resulting black hole has a smaller entropy than the original 
black hole, this relaxation process is unitary because $k'$ in the 
left-hand side runs only over $1,\cdots,e^{S_0(M)}/2 = e^{S_0(M-E_1)}$. 
We note that the creation of a positive energy Hawking quantum 
and a negative energy excitation in Eq.~(\ref{eq:H-toy-1}) (and 
in Eq.~(\ref{eq:H-emission})) takes a form very different from the 
standard ``pair creation'' of particles, which is often invoked to 
visualize the Hawking emission process.  In the pair creation picture, 
the positive and negative energy excitations are maximally entangled 
with each other, which is not the case here.  In fact, it is this 
lack of entanglement that allows the emission process to transfer 
the information from the black hole to radiation.

We emphasize that from the semiclassical spacetime viewpoint, the emission 
of Eq.~(\ref{eq:H-emission}) is viewed as occurring locally around the 
edge of the zone, which is possible because the information about the 
black hole microstate extends into the whole zone region according to 
Eqs.~(\ref{eq:Psi_k-decomp},~\ref{eq:distr}).  To elucidate this point, 
we may consider the tortoise coordinate
\begin{equation}
  r^* = r + 2M l_{\rm P}^2\, \ln \frac{r-2M l_{\rm P}^2}{2M l_{\rm P}^2},
\label{eq:tortoise}
\end{equation}
in which the region outside the Schwarzschild horizon $r \in (2M l_{\rm P}^2, 
\infty)$ is mapped into $r^* \in (-\infty,\infty)$.  This coordinate 
is useful in that the kinetic term of an appropriately redefined 
field takes the canonical form, so that its propagation can be 
analyzed as in flat space.  In this coordinate, the stretched 
horizon, located at $r = 2M l_{\rm P}^2 + O(l_*^2/M l_{\rm P}^2)$ 
(see footnote~\ref{ft:stretched}), is at
\begin{equation}
  r^*_{\rm s} \simeq -4M l_{\rm P}^2 \ln\frac{M l_{\rm P}^2}{l_*} 
  \simeq -4M l_{\rm P}^2 \ln(M l_{\rm P}),
\label{eq:rs_stretched}
\end{equation}
where $l_*$ is the string (or gravitational cutoff) scale, which we 
take to be within a couple of orders of magnitude of $l_{\rm P}$.  This 
implies that there is a large distance between the stretched horizon and 
the potential barrier region when measured in $r^*$:\ $\varDelta r^* 
\approx 4M l_{\rm P}^2 \ln(M l_{\rm P}) \gg O(M l_{\rm P}^2)$ for 
$\ln(M l_{\rm P}) \gg 1$.  On the other hand, a localized Hawking 
quantum is represented by a wavepacket with width of $O(M l_{\rm P}^2)$ 
in $r^*$, since it has an energy of order $T_{\rm H} = 1/8\pi M 
l_{\rm P}^2$ defined in the asymptotic region.

The point is that, given the state $\ket{\Psi_k(M)} = \ket{\psi_k(M)} 
\ket{\phi_I}$, the process in Eq.~(\ref{eq:H-emission}) occurs in the 
region $|r^*| \approx O(M l_{\rm P}^2)$ (i.e.\ the region in which the 
effective gravitational potential starts shutting off toward large $r^*$) 
without involving deep interior of the zone $-r^* \gg M l_{\rm P}^2$. 
In this region, information stored in the vacuum state is converted 
into that of a particle state outside the zone.  More specifically, the 
information in the vacuum represented by the $k$ index (which may also 
be viewed as a thermal bath of infrared modes, Eq.~(\ref{eq:IR-modes}), 
though only in certain senses) is transferred into that in modes 
$a_{\rm far} \neq 0$, i.e.\ Hawking quanta, which have clear independent 
identities over the background spacetime.  Due to energy conservation, 
this process is accompanied by the creation of ingoing negative 
energy excitations; however, they are not maximally entangled with 
the emitted Hawking quanta.

\begin{figure}[t]
\begin{center}
  \includegraphics[height=10cm]{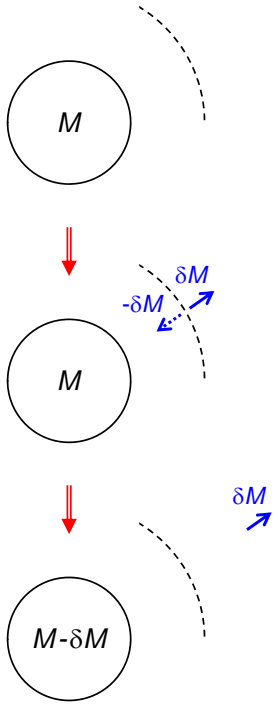}
\end{center}
\caption{A schematic picture of the elementary Hawking emission process; 
 time flows from the top to the bottom.  The edge of the zone, i.e.\ the 
 barrier region of the effective gravitational potential, is shown by a 
 portion of a dashed circle at each moment in time.  The emitted Hawking 
 quanta as well as negative energy excitations are depicted by arrows 
 (solid and dotted, respectively) although they are mostly $s$-waves.}
\label{fig:emission}
\end{figure}
In Fig.~\ref{fig:emission}, we depict schematically the elementary 
Hawking emission process described here.  In the figure, we have denoted 
the emitted Hawking quanta as well as negative energy excitations by 
arrows, although they are mostly $s$-waves~\cite{Page:1976df}.  The 
discussion here makes it clear that the purifiers of the emitted Hawking 
quanta in the Hawking emission process are microstates which semiclassical 
theory describes as a vacuum.  In particular, the emission process does 
{\it not} involve any excitation which, in the near horizon Rindler 
approximation, appears as a mode breaking entanglement between the two 
Rindler wedges necessary to keep the horizon smooth.  Outgoing Hawking 
quanta emerge at the edge of the zone, living outside the applicability 
of the Rindler approximation.  Ingoing negative energy excitations 
appear, in the Rindler approximation, as modes smooth in Minkowski space, 
which involve necessary entanglements between Rindler modes in the two 
wedges and have frequencies of order $1/M l_{\rm P}^2$ in the Minkowski 
frame.  Unlike what was considered in Ref.~\cite{Almheiri:2012rt}, and 
unlike what a ``naive'' interpretation of semiclassical theory might seem 
to suggest, Hawking quanta are not modes associated solely with one of the 
Rindler wedges ($b$ modes in the notation of Ref.~\cite{Almheiri:2012rt}) 
nor outgoing Minkowski modes ($a$ modes), which would appear to have 
high energies for observers who are freely falling into the black 
hole.  This allows for avoiding the entropy argument for firewalls 
given in Ref.~\cite{Almheiri:2012rt} as well as the typicality argument 
in Ref.~\cite{Marolf:2013dba}.

In the discussion of the Hawking emission so far, we have assumed that 
a single emission of Hawking quanta as well as the associated creation 
of ingoing negative energy excitations occur in a black hole vacuum 
state consisting of $\ket{\Psi_k(M)}$'s, which are defined in the limit 
that the evolution effect is ignored.  In reality, however, there are 
always of order $\ln (M l_{\rm P})$ much of negative energy excitations 
in the zone, since the emission process occurs in every time interval 
of order $M l_{\rm P}^2$ and the time it takes for a negative energy 
excitation to reach the stretched horizon is of order $M l_{\rm P}^2 
\ln (M l_{\rm P})$ (both measured in the asymptotic region)---an 
evaporating black hole has an ingoing flux of negative energy excitations 
of entropy $\approx O(-\ln (M l_{\rm P}))$ at all times.  This flux of 
excitations modifies spacetime geometry from that of a Schwarzschild 
black hole; in particular, the geometry near the horizon is well 
described by the advanced/ingoing Vaidya metric~\cite{Bardeen:1981zz}. 
Note that as discussed in Section~\ref{subsec:semiclassical}, we may 
redefine our vacuum states to include these negative energy excitations, 
although we do not do it here.

\begin{figure}[t]
\begin{center}
  \subfigure[Time reversal of Fig.~\ref{fig:emission}]{\includegraphics[height=10cm]{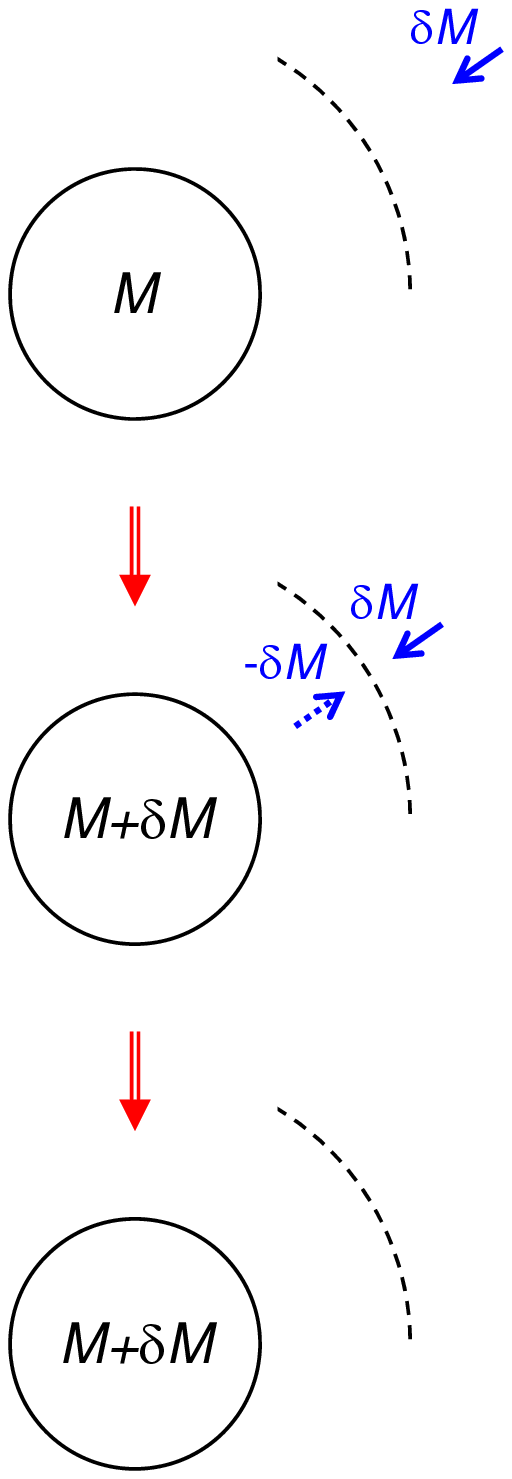}}
\hspace{3cm}
  \subfigure[Generic incoming radiation]{\includegraphics[height=10cm]{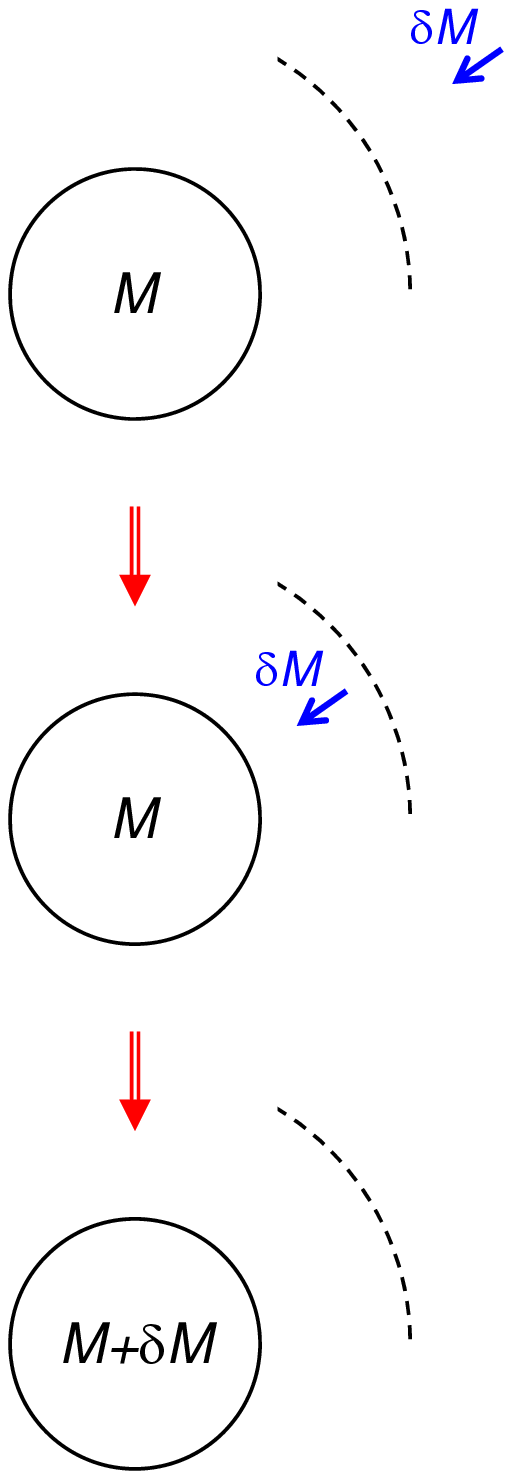}}
\end{center}
\caption{Time reversal of the Hawking emission process (a) as opposed 
 to the process in which generic incoming radiation enters into the zone 
 of a usual black hole (b).  The former is an entropy decreasing process 
 requiring an exponentially fine-tuned initial state, while the latter 
 is a standard process respecting the (generalized) second law of 
 thermodynamics.}
\label{fig:reverse}
\end{figure}
Finally, it is instructive to consider the time reversal of the Hawking 
emission process.  In this case, radiation coming from the far exterior 
region and outgoing negative energy excitations emitted from the stretched 
horizon meet around the edge of the zone; see Fig.~\ref{fig:reverse}(a). 
This results in a black hole state of mass given by the sum of the mass 
$M$ of the original black hole (before emitting the negative energy 
excitations) and the energy $\delta M$ of the incoming radiation.  It 
is a ``vacuum'' state in the sense that there is no excitation in the 
zone except for those associated with a steady flux of outgoing negative 
energy excitations.  We emphasize that this process is very different 
from what happens when generic incoming radiation of energy $\delta M 
\approx O(1/M l_{\rm P}^2)$ is sent to a usual (i.e.\ evaporating, 
not anti-evaporating) black hole.  In this case, the radiation enters 
into the zone without being ``annihilated'' by a negative energy 
excitation, which after hitting the stretched horizon will lead to 
a black hole state of mass $M + \delta M$; see Fig.~\ref{fig:reverse}(b). 
In fact, the process in Fig.~\ref{fig:reverse}(a) is a process which 
leads to a decrease of coarse-grained (or thermal) entropy, as implied 
by the fact that the coarse-grained entropy increases in the standard 
Hawking emission process~\cite{Zurek:1982zz}.  In order for this to 
happen, therefore, the initial radiation and black hole state must be 
exponentially fine-tuned; otherwise, the radiation would simply propagate 
inward in the zone as depicted in Fig.~\ref{fig:reverse}(b) (although 
it can be subject to significant scattering by the effective gravitational 
potential at the time of the entrance). The origin of the conversion 
from radiation to vacuum degrees of freedom for such a fine-tuned 
initial state can be traced to the non-decoupling of the $a$ and $k$ 
indices discussed in Section~\ref{subsec:semiclassical}.%
\footnote{If the black hole vacuum states are redefined as discussed 
 in Section~\ref{subsec:semiclassical}, the outgoing negative energy 
 flux cannot be seen as excitations.  The physics described here, however, 
 will not change; in particular, only exponentially fine-tuned initial 
 states allow for converting radiation to vacuum degrees of freedom 
 around the edge of the zone.}

\subsubsection{Semiclassical (thermal) description}

The expression in Eq.~(\ref{eq:BH-evol}) implies that at an intermediate 
stage of the evolution, the information about the initial collapsing 
matter is encoded in the black hole microstates labeled by $k$ and their 
entanglement with the rest of the system (which will later be transformed 
into the state of final-state Hawking radiation).  Since semiclassical 
theory is incapable of describing the dynamics associated with the index 
$k$, it leads to apparent violation of unitarity {\it at all stages} 
of the black hole formation and evaporation processes.  In particular, 
the state of the emitted Hawking quanta in each time interval of order 
$M(t) l_{\rm P}^2$ is given by the incoherent thermal superposition with 
temperature $1/8\pi M(t) l_{\rm P}^2$, making the final Hawking radiation 
state a mixed thermal state---this is an intrinsic limitation of the 
semiclassical description, which involves a coarse-graining.

To see in detail how thermal Hawking radiation in the semiclassical 
picture results from unitary evolution at the fundamental level, 
let us analyze the elementary Hawking emission process given in 
Eq.~(\ref{eq:H-process}).  Following Eq.~(\ref{eq:rho_0}), we consider 
the ``semiclassical vacuum state'' with a black hole of mass $M$, 
obtained after taking the maximally mixed ensemble of microstates:
\begin{equation}
  \rho(M) = \frac{1}{e^{S_0(M)}} \sum_{k=1}^{e^{S_0(M)}} 
    \ket{\psi_k(M)} \ket{\phi_I} \bra{\psi_k(M)} \bra{\phi_I}.
\label{eq:sc-rho_before}
\end{equation}
The evolution of this state under Eq.~(\ref{eq:H-process}) is then given by
\begin{equation}
  \rho(M) \rightarrow 
    \frac{1}{e^{S_0(M)}} \sum_{k=1}^{e^{S_0(M)}}\; \sum_{i,i'} 
    \sum_{k_i = 1}^{e^{S_0(M-E_i)}} \sum_{k'_{i'} = 1}^{e^{S_0(M-E_{i'})}} 
    \alpha^k_{i k_i} \alpha^{k*}_{i' k'_{i'}} 
    \ket{\psi_{k_i}(M-E_i)} \ket{\phi_{I+i}} 
    \bra{\psi_{k'_{i'}}(M-E_{i'})} \bra{\phi_{I+i'}}.
\label{eq:emission-mixed}
\end{equation}
Now, assuming that the microscopic dynamics of the vacuum degrees of 
freedom are generic, we expect using $S_0(M) = 4\pi M^2 l_{\rm P}^2$ 
that tracing out the black hole states leads to
\begin{equation}
  {\rm Tr}\, \biggl[ 
    \frac{1}{e^{S_0(M)}} \sum_{k=1}^{e^{S_0(M)}} 
    \sum_{k_i = 1}^{e^{S_0(M-E_i)}} \sum_{k'_{i'} = 1}^{e^{S_0(M-E_{i'})}} 
    \alpha^k_{i k_i} \alpha^{k*}_{i' k'_{i'}} 
    \ket{\psi_{k_i}(M-E_i)} \bra{\psi_{k'_{i'}}(M-E_{i'})} \biggr] 
  \approx \frac{1}{Z} g_i e^{-\frac{E_i}{T_{\rm H}}} \delta_{ii'},
\label{eq:sc-trace}
\end{equation}
where $T_{\rm H} = 1/8\pi M l_{\rm P}^2$, $Z = \sum_i g_i e^{-E_i/T_{\rm H}}$, 
and $g_i$ is a factor that depends on $i$.  This allows us to write the 
reduced density matrix representing the exterior state after the evolution 
in Eq.~(\ref{eq:emission-mixed}) as
\begin{equation}
  \rho_{\rm ext} \approx \frac{1}{Z} \sum_i 
    g_i e^{-\frac{E_i}{T_{\rm H}}} \ket{\phi_{I+i}} \bra{\phi_{I+i}},
\label{eq:sc-final}
\end{equation}
which is the result obtained in Hawking's original calculation, with 
$g_i$ representing the gray-body factor calculable in the semiclassical 
theory~\cite{Page:1976df}.

The analysis given above elucidates why the semiclassical calculation 
sees apparent violation of unitarity in the Hawking emission process, 
i.e.\ why the final expression in Eq.~(\ref{eq:sc-final}) does not 
depend on microstates of the black hole, despite the fact that the 
elementary process in Eq.~(\ref{eq:H-process}) is unitary, so that 
the coefficients $\alpha^k_{i k_i}$ depend on $k$.  It is because 
the semiclassical calculation (secretly) deals with the mixed state, 
Eq.~(\ref{eq:sc-rho_before}), from the beginning---states in semiclassical 
theory are maximal mixtures of black hole microstates labeled by vacuum 
indices, i.e.\ $k$'s.  By construction, the semiclassical theory cannot 
capture unitarity of detailed microscopic processes involving these 
indices, including the black hole formation and evaporation processes.

We finally discuss how the unitarity and thermal nature of the black 
hole evaporation process may appear in (thought) experiments, illuminating 
physical implications of the picture described here.  Suppose we prepare 
an ensemble of a large number of black holes of mass $M$ all of which 
are in an identical microstate $k$, and collect the Hawking quanta 
emitted from these black holes in a time interval of order $M l_{\rm P}^2$. 
The quanta emitted from each black hole are then in the same quantum state 
throughout the ensemble, so that a measurement of the spectrum of all 
the emitted quanta does {\it not} reveal the thermal property predicted 
by the semiclassical theory.  On the other hand, if the members of the 
ensemble are in different microstates distributed randomly in $k$ space, 
then the collection of the Hawking quanta emitted from all the black 
holes do exhibit the thermal nature consistent with the prediction 
of the semiclassical theory within the Hilbert space describing the 
quanta emitted from {\it each} black hole (which has dimension only 
of order unity).

What is the significance of the thermal nature for a single black hole, 
rather than an ensemble of a large number of black holes?  If we form 
a black hole of mass $M$ in a particular microstate $k$ and collect 
all the Hawking quanta emitted throughout the evaporation process 
{\it without measuring them along the way}, then the state of the 
quanta contains the complete information about $k$, reflecting unitarity 
of the process at the fundamental level---the concept of thermality 
does not apply to this particular state {\it as a whole}.  On the 
other hand, if an observer measures Hawking quanta emitted in each 
time interval of order $M(t) l_{\rm P}^2$, then the (incoherent) 
ensemble of measurement outcomes does exhibit the thermal nature 
as predicted by the semiclassical theory.%
\footnote{In the more fundamental, many-world picture, this implies 
 that the record of a physical observer who has ``measured,'' or 
 interacted with, emitted quanta in multiple moments shows a result 
 consistent with the thermality predicted by the semiclassical theory. 
 Note that a single branch in which such an observer lives does {\it not} 
 in general contain the whole information about the initial black hole 
 state $k$.  The complete information about $k$ (as well as that of 
 the initial state of the observer) is contained only in a state given 
 by a superposition of all possible branches resulting from interactions 
 (and non-interactions) between the observer and quanta, representing 
 all the possible ``outcomes'' the observer could have had (the 
 probability distribution of which is consistent with thermality).}
Since this is the kind of measurement that a realistic observer typically 
makes, the semiclassical theory can be said to provide a good prediction 
even for the outcome of (a series of) measurements a single observer 
performs on a single black hole.

\subsection{Black hole mining---``microscopic'' and semiclassical 
 descriptions}
\label{subsec:mining}

It is known that one can accelerate the energy loss rate of a black hole 
faster than that of spontaneous Hawking emission by extracting its energy 
from the thermal atmosphere using a physical apparatus:\ the mining 
process.  This acceleration occurs largely because the number of 
``channels'' one can access increases by going into the zone---unlike 
the case of spontaneous Hawking emission, which is dominated by $s$-wave 
radiation, higher angular momentum modes can also contribute to the 
energy loss in this process~\cite{Brown:2012un}.  Note that the rate 
of energy loss associated with {\it each} channel, however, is still 
the same order as that in the spontaneous Hawking emission process:\ 
energy of order $E \approx O(1/M l_{\rm P}^2)$ is lost in each time 
interval of $t \approx O(M l_{\rm P}^2)$, with $E$ and $t$ both 
defined in the asymptotic region.  This fact will become important 
in Section~\ref{sec:infalling} when we discuss the mining process 
as viewed from an infalling reference frame.

The information transfer associated with the mining process occurs 
in a similar way to that in the spontaneous Hawking emission process. 
An essential difference is that since the process involves higher 
angular momentum modes, the negative energy excitations arising from 
backreactions can now be localized in angular directions.  Specifically, 
consider a physical detector (or a system of detectors) located at 
a fixed Schwarzschild radial coordinate $r = r_{\rm d}$ within the 
zone, $r_{\rm s} < r_{\rm d} < R_{\rm Z}$.  The detector then responds 
as if it is immersed in the thermal bath of blueshifted Hawking temperature 
$T(r_{\rm d})$, with $T(r)$ given by Eq.~(\ref{eq:T_local}).  Suppose 
the detector has the ground state $\ket{d_0}$ and excited states 
$\ket{d_i}$ ($i=1,2,\dots$) playing the role of the ``ready'' state 
and pointer states, respectively, and that the proper energies needed 
to excite $\ket{d_0}$ to $\ket{d_i}$ are given by $E_{{\rm d},i}$. 
The mining process can then be written such that after a timescale 
of $t \approx O(M l_{\rm P}^2)$ (as measured in the asymptotic region), 
the state of the combined black hole and detector system evolves as
\begin{equation}
  \ket{\psi_k(M)} \ket{d_0} \rightarrow 
    \sum_{i,a,k'} c^k_{i a k'} \ket{\psi_{a; k'}(M)} \ket{d_i},
\label{eq:mining-1}
\end{equation}
where we have assumed, as in the discussion of ``elementary'' Hawking 
emission, that there are no excitations other than those directly 
associated with the process.  The state $\ket{\psi_{a; k'}(M)}$ arises 
as a result of backreaction of the detector response; it contains a 
negative energy excitation $a$ with energy $-E_a$, which is generally 
localized in angular directions.  The coefficients $c^k_{i a k'}$ are 
nonzero only if $E_a \approx E_{{\rm d},i} \sqrt{1-2Ml_{\rm P}^2/r_{\rm d}}$ 
within the uncertainty.

Once created, the negative energy excitations propagate inward, and 
after time of $t \approx r^*_{\rm d} - r^*_{\rm s}$ collide with 
the stretched horizon, where $r^*$ is the tortoise coordinate in 
Eq.~(\ref{eq:tortoise}).  This will make the black hole states 
relax as
\begin{equation}
  \ket{\psi_{a; k'}(M)} \rightarrow 
    \sum_{k_a} d^{a k'}_{k_a} \ket{\psi_{k_a}(M-E_a)},
\label{eq:mining-2}
\end{equation}
in the scrambling time of $t \approx O(M l_{\rm P}^2 \ln (M l_{\rm P}))$. 
As in the case of spontaneous Hawking emission, this relaxation 
process is unitary because the negative energy excitations carry 
negative entropies; i.e.\ for a fixed $a$, the index $k'$ runs only 
over $1,\cdots,e^{S_0(M-E_a)} \ll e^{S_0(M)}$.  The combination of 
Eqs.~(\ref{eq:mining-1},~\ref{eq:mining-2}) then yields
\begin{equation}
  \ket{\psi_k(M)} \ket{d_0} \rightarrow 
    \sum_{i,k_i} \alpha^k_{i k_i} \ket{\psi_{k_i}(M-E_i)} \ket{d_i},
\label{eq:mining-3}
\end{equation}
where $\alpha^k_{i k_i} = \sum_{a,k'} c^k_{i a k'} d^{a k'}_{k_i}$ and 
$E_i = E_{{\rm d},i} \sqrt{1-2Ml_{\rm P}^2/r_{\rm d}}$.  This represents 
a microscopic, unitary description of the elementary mining process.

In the description given above, we have separated the detector state 
from the state of the black hole, but in a treatment fully consistent 
with the notation in earlier sections, the detector itself must be viewed 
as excitations over $\ket{\psi_k(M)}$.  After the detector response 
process in Eq.~(\ref{eq:mining-1}), these excitations can be entangled 
with Hawking quanta emitted earlier, reflecting the fact that the detector 
can extract information from the black hole.  Since the detector can 
now be put deep in the zone, in which the Rindler approximation is 
applicable, this implies that excitations localized within the Rindler 
wedge corresponding to the region $r > r_{\rm s}$ are entangled with 
early Hawking radiation.  Does this lead to firewalls as discussed 
in Ref.~\cite{Almheiri:2012rt}?  The answer is no.  The excitations 
describing the detector are, in the near horizon Rindler approximation, 
those of modes that are smooth in Minkowski space ($a$ modes in the 
notation of Ref.~\cite{Almheiri:2012rt}).  Likewise, modes representing 
negative energy excitations arising from the backreactions are also 
ones smooth in Minkowski space.  Excitations of these modes, of course, 
{\it do} perturb the black hole system, which can indeed be significant 
if the detector is held very close to the horizon.  This effect, however, 
is caused by physical interactions between the detector and vacuum 
degrees of freedom, and is confined in the causal future of the 
interaction event.  This is not the firewall phenomenon.

The semiclassical description of the mining process in 
Eq.~(\ref{eq:mining-3}) is obtained by taking maximal mixture for 
the vacuum indices.  Specifically, the semiclassical state before 
the process starts is given by
\begin{equation}
  \rho(M) = \frac{1}{e^{S_0(M)}} \sum_{k=1}^{e^{S_0(M)}} 
    \ket{\psi_k(M)} \ket{d_0} \bra{\psi_k(M)} \bra{d_0}.
\label{eq:sc_mining-1}
\end{equation}
The evolution of this state under Eq.~(\ref{eq:mining-3}) is then
\begin{equation}
  \rho(M) \rightarrow 
    \frac{1}{e^{S_0(M)}} \sum_{k=1}^{e^{S_0(M)}}\; \sum_{i,i'} 
    \sum_{k_i = 1}^{e^{S_0(M-E_i)}} \sum_{k'_{i'} = 1}^{e^{S_0(M-E_{i'})}} 
    \alpha^k_{i k_i} \alpha^{k*}_{i' k'_{i'}} 
    \ket{\psi_{k_i}(M-E_i)} \ket{d_i} 
    \bra{\psi_{k'_{i'}}(M-E_{i'})} \bra{d_{i'}}.
\label{eq:sc_mining-2}
\end{equation}
This leads to the density matrix describing the detector state after 
the process
\begin{equation}
  \rho_{\rm d} = \sum_{i,i'} \gamma_{ii'}\, \ket{d_i} \bra{d_{i'}},
\label{eq:rho_d}
\end{equation}
where
\begin{equation}
  \gamma_{ii'} = {\rm Tr}\, \biggl[ 
    \frac{1}{e^{S_0(M)}} \sum_{k=1}^{e^{S_0(M)}} 
    \sum_{k_i = 1}^{e^{S_0(M-E_i)}} \sum_{k'_{i'} = 1}^{e^{S_0(M-E_{i'})}} 
    \alpha^k_{i k_i} \alpha^{k*}_{i' k'_{i'}} 
    \ket{\psi_{k_i}(M-E_i)} \bra{\psi_{k'_{i'}}(M-E_{i'})} \biggr].
\label{eq:gamma_iip}
\end{equation}
Assuming that the microscopic dynamics of the vacuum degrees of freedom 
are generic, $\gamma_{ii'}$ is expected to take the form
\begin{equation}
  \gamma_{ii'} \approx \frac{1}{Z} f_i 
    e^{-\frac{E_{{\rm d},i}}{T(r_{\rm d})}}\, \delta_{ii'},
\label{eq:gamma-sc}
\end{equation}
where $Z = \sum_i f_i e^{-E_{{\rm d},i}/T(r_{\rm d})}$, and $f_i$ is 
the detector response function reflecting intrinsic properties of the 
detector under consideration.  This implies that in the semiclassical 
approximation, the final detector state does not have any information 
about the original black hole microstate, despite the fact that the 
fundamental process in Eq.~(\ref{eq:mining-3}) is, in fact, unitary.

\subsection{The fate of an infalling object}
\label{subsec:obj}

We now discuss how an object falling into a black hole is described 
in a distant reference frame.  As we have seen, having a well-defined 
black hole geometry requires a superposition of an enormous number of 
energy-momentum eigenstates.  While the necessary spreads in energy 
and momentum are small when measured in the asymptotic region, the 
spreads of {\it local} energy and momentum (i.e.\ those measured by 
local approximately static observers) are large in the region close 
to the horizon, because of large gravitational blueshifts.  This makes 
the local temperature $T(r)$ associated with the vacuum degrees of 
freedom, Eq.~(\ref{eq:T_local}), very high near the horizon.  We expect 
that the semiclassical description becomes invalid when this temperature 
exceeds the string (cutoff) scale, $T(r) \gtrsim 1/l_*$.  Namely, 
semiclassical spacetime exists only in the region
\begin{equation}
  r > r_{\rm s} 
  = 2Ml_{\rm P}^2 + O\biggl( \frac{l_*^2}{M l_{\rm P}^2} \biggr),
\label{eq:stretched}
\end{equation}
where $r_{\rm s}$ is identified as the location of the stretched horizon. 
The same conclusion can also be obtained by demanding that the gravitational 
thermal entropy stored in the region where the semiclassical spacetime 
picture is applicable is a half of the Bekenstein-Hawking entropy, 
${\cal A}/8 l_{\rm P}^2$, as discussed in footnote~\ref{ft:stretched}.

Let us consider that an object is dropped from $r = r_0$ with vanishing 
initial velocity, where $r_0 - 2Ml_{\rm P}^2 \approx O(M l_{\rm P}^2) > 0$. 
It then freely falls toward the black hole and hits the stretched horizon 
at $r = r_{\rm s}$ in Schwarzschild time of about $4 M l_{\rm P}^2 
\ln(M l_{\rm P}^2/l_*)$.  Before it hits the stretched horizon, the 
object is described by $a$ and $a_{\rm far}$, the indices labeling 
field and string theoretic excitations over the semiclassical background 
spacetime.  After hitting the stretched horizon, the information about 
the object will move to the index $\bar{a}$, labeling excitations of 
the stretched horizon.  The information about the fallen object will 
then stay there, at least, for the thermalization (or scrambling) time 
of the stretched horizon, of order $M l_{\rm P}^2 \ln(M l_{\rm P})$. 
This allows for avoiding the inconsistency of quantum cloning in black 
hole physics~\cite{Hayden:2007cs}.  Finally, the information in $\bar{a}$ 
will further move to $k$, which can (later) be extracted by an observer 
in the asymptotic region via the Hawking emission or mining process, 
as described in the previous two subsections.

We note that the statement that an object is in the semiclassical regime 
(i.e.\ represented by indices $a$ and $a_{\rm far}$) does {\it not} 
necessarily mean that it is well described by semiclassical {\it field} 
theory.  Specifically, it is possible that stringy effects become important 
before the object hits the stretched horizon.  As an example, consider 
dropping an elementary particle of mass $m$ ($\ll 1/l_*$) from $r=r_0$ 
with zero initial velocity.  (Here, by elementary we mean that there 
is no composite structure at lengthscale larger than $l_*$.)  The local 
energy and local radial momentum of the object will then vary, as it 
falls, as:
\begin{equation}
  E_{\rm loc} = m \sqrt{\frac{1-\frac{2Ml_{\rm P}^2}{r_0}} 
    {1-\frac{2Ml_{\rm P}^2}{r}}},
\qquad\quad
  p_{\rm loc} = -m 
    \sqrt{\frac{\frac{2Ml_{\rm P}^2}{r}-\frac{2Ml_{\rm P}^2}{r_0}} 
    {1-\frac{2Ml_{\rm P}^2}{r}}}.
\label{eq:Ep_obj}
\end{equation}
The values of $E_{\rm loc} \approx -p_{\rm loc}$ get larger as $r$ gets 
smaller, and for $m \gg 1/M l_{\rm P}^2$ (which we assume here) become 
of order $1/l_*$ before the object hits the stretched horizon, i.e.\ at
\begin{equation}
  r - 2M l_{\rm P}^2 \simeq 2M l_{\rm P}^2 (m l_*)^2 
    \biggl( 1 - \frac{2 M l_{\rm P}^2}{r_0} \biggr).
\label{eq:stringy}
\end{equation}
The Schwarzschild time it takes for the object to reach this point is only 
about $-4 M l_{\rm P}^2 \ln(m l_*)$, much smaller than the time needed 
to reach the stretched horizon, $4 M l_{\rm P}^2 \ln(M l_{\rm P}^2/l_*)$. 
After the object reaches this point, i.e.\ when $E_{\rm loc} \approx 
-p_{\rm loc} \gtrsim 1/l_*$, stringy effects might become important; 
specifically, its Lorentz contraction saturates and transverse size grows 
with $E_{\rm loc}$~\cite{Susskind:1993aa}.  Note that this dependence 
of the description on the boost of a particle does not necessarily 
mean violation of Lorentz invariance---physics can still be fully 
Lorentz invariant.%
\footnote{It is illuminating to consider how these stringy effects appear 
 in a two-particle scattering process in Minkowski space.  For $\sqrt{s} 
 \lesssim 1/l_*$, where $s$ is the Mandelstam variable, there is 
 a reference frame in which energies/momenta of {\it both} particles 
 are smaller than $1/l_*$, guaranteeing that these effects are not 
 important in the process.  For $\sqrt{s} > 1/l_*$, on the other hand, 
 at least one particle has an energy/momentum larger than $1/l_*$ in 
 {\it any} reference frame, suggesting that stringy effects become 
 important in scattering with such high $\sqrt{s}$.}

\begin{figure}[t]
\begin{center}
  \includegraphics[width=13.5cm]{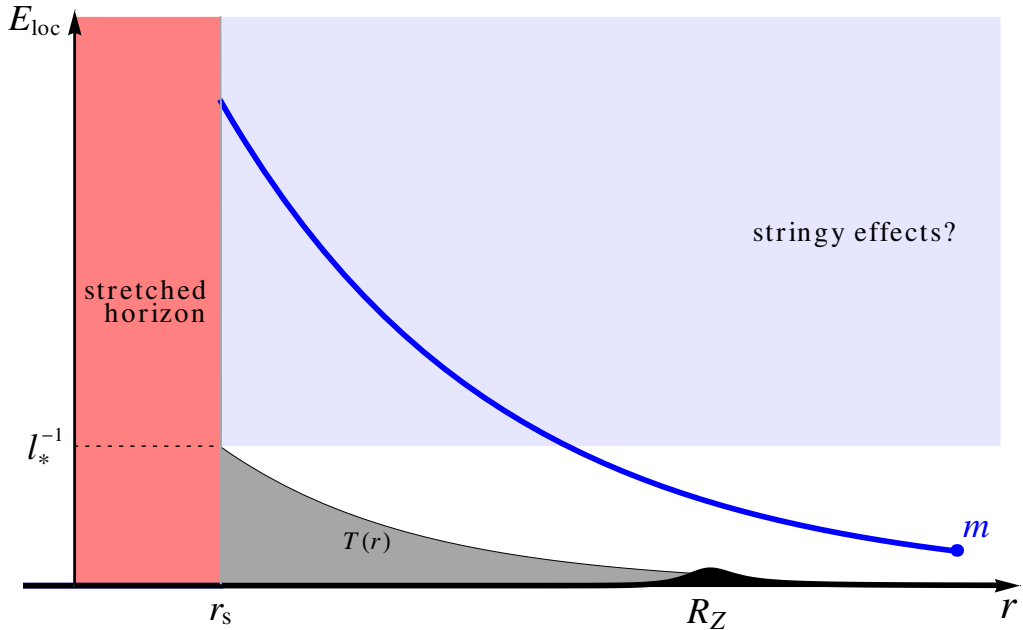}
\end{center}
\caption{A schematic depiction of the fate of an elementary particle 
 of mass $m$ ($1/M l_{\rm P}^2 \ll m \ll 1/l_*$) dropped into a black 
 hole, viewed in a distant reference frame.  As the particle falls, its 
 local energy blueshifts and exceeds the string/cutoff scale $1/l_*$ 
 before it hits the stretched horizon.  After this point, stringy 
 effects could become important, although the semiclassical description 
 of the object may still be applicable.  The object hits the stretched 
 horizon at a Schwarzschild time of about $4 M l_{\rm P}^2 \ln(M 
 l_{\rm P}^2/l_*)$ after the drop.  After this time, the semiclassical 
 description of the object is no longer applicable, and the information 
 about the object will be encoded in the index $\bar{a}$, representing 
 excitations of the stretched horizon.  (This information will further 
 move to the vacuum index $k$ later, so that it can be extracted by 
 an observer in the asymptotic region via the Hawking emission or 
 mining process.)}
\label{fig:falling}
\end{figure}
A schematic picture for the fate of an infalling object described above 
is given in Fig.~\ref{fig:falling}.  In a distant reference frame, the 
semiclassical description of the object is applicable only until it hits 
the stretched horizon, after which it is represented as excitations of 
the stretched horizon.  On the other hand, according to general relativity 
(or the equivalence principle), the falling object does not experience 
anything other than smooth empty spacetime when it crosses the horizon, 
except for effects associated with curvature, which are very small for 
a black hole of mass $M \gg 1/l_{\rm P}$.  If this picture is correct, 
then we expect there is a way to reorganize the dynamics of the stretched 
horizon such that the general relativistic smooth interior of the black 
hole becomes manifest.  In the complementarity picture, this is achieved 
by performing an appropriate reference frame change.  We now move on 
to discuss this issue.

\section{Black Hole---An Infalling Description}
\label{sec:infalling}

In order to describe the fate of an infalling object using low energy 
language after it crosses the Schwarzschild horizon, we need to perform 
a change of the reference frame from a distant one, which we have been 
considering so far, to an infalling one which falls into the black hole 
with the object.  In general, studying this issue is complicated by the 
fact that the general and precise formulation of complementarity is not 
yet known, but we may still explore the expected physical picture based 
on some general considerations.

The aim of this section is to argue that the existence of interior 
spacetime, as suggested by general relativity, does not contradict 
the unitarity of the Hawking emission and black hole mining processes, 
as described in the previous section in a distant reference frame. 
We do this by first arguing that there exists a reference frame---an 
infalling reference frame---in which the spacetime around a point on 
the Schwarzschild horizon appears as a large nearly flat region, with 
the curvature lengthscale of order $M l_{\rm P}^2$.  This is a reference 
frame whose origin falls freely from rest from a point sufficiently 
far from the black hole.  We discuss how the description based on 
this reference frame is consistent with that in the distant reference 
frame, despite the fact that they apparently look very different, 
for example in spacetime locations of the vacuum degrees of freedom.

We then discuss how the system is described in more general reference 
frames, in particular a reference frame whose origin falls from 
rest from a point close to the Schwarzschild horizon.  We will also 
discuss (non-)relations of black hole mining by a near-horizon static 
detector and the---seemingly similar---Unruh effect in Minkowski space. 
The discussion in this section illuminates how general coordinate 
transformations may work at the level of full quantum gravity, beyond 
the approximation of quantum field theory in curved spacetime.

\subsection{Emergence of interior spacetime---free fall from a distance}
\label{subsec:interior}

What does a reference frame really mean?  According to the general 
complementarity picture described in Section~\ref{sec:failure}, it 
corresponds to a foliation of a portion of spacetime which a single 
(hypothetical) observer can access.  As discussed there, the procedure 
to erect such a reference frame should not depend on the background 
geometry in order for the framework to be applicable generally, and 
there is currently no precise, established formulation to do that 
(although there are some partially successful attempts; see, e.g., 
Ref.~\cite{Nomura:2013nya}).  Here we focus only on classes of reference 
frames describing the same system with a fixed black hole background. 
This limitation allows us to bypass many of the issues arising when 
we consider the most general application of the complementarity picture.

In this subsection, we consider a class of reference frames which 
we call infalling reference frames.  We argue that a reference frame 
in this class makes it manifest that the spacetime near the origin of 
the reference frame appears as a large approximately flat region when 
it crosses the Schwarzschild horizon, up to corrections from curvature 
of lengthscale $M l_{\rm P}^2$.  We discuss how the interior spacetime 
of the black hole can emerge through the complementarity transformation 
representing a change of reference frame from the distant to infalling 
ones.  Consistency of the infalling picture described here with the 
distant frame description in Section~\ref{sec:distant} will be discussed 
in more detail in the next subsection.

We consider a reference frame associated with a freely falling (local 
Lorentz) frame, with its spatial origin $p_0$ following the worldline 
representing a hypothetical observer~\cite{Nomura:2011rb,Nomura:2013nya}. 
In particular, we let the origin of the reference frame, $p_0$, follow 
the trajectory of a timelike geodesic, representing the observer who 
is released from rest at $r = r_0$, with $r_0$ sufficiently far from 
the Schwarzschild horizon, $r_0 - 2 M l_{\rm P}^2 \gtrsim M l_{\rm P}^2$. 
According to the complementarity hypothesis, the system described in this 
reference frame does not have a (hot) stretched horizon at the location 
of the Schwarzschild horizon when $p_0$ crosses it.  (The stretched 
horizon must have existed around the Schwarzschild horizon when $p_0$ 
was far away, $r_{p_0} - 2 M l_{\rm P}^2 \gtrsim O(M l_{\rm P}^2)$, 
because the description in those earlier times must be approximately 
that of a distant reference frame, i.e.\ that discussed in the 
previous section.)  In particular, the region around $p_0$ must 
appear approximately flat, i.e.\ up to small effects from curvature 
of order $1/M^2 l_{\rm P}^4$, until $p_0$ approaches the singularity.

\begin{figure}[t]
\begin{center}
  \includegraphics[width=14cm]{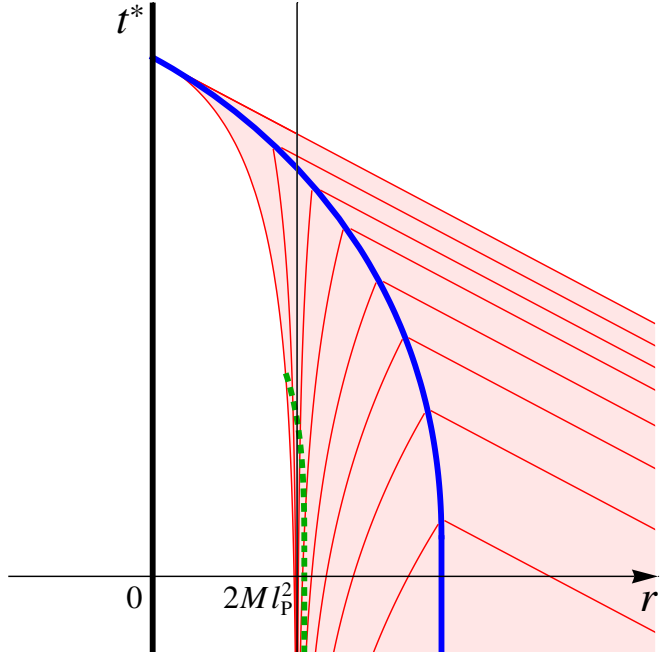}
\end{center}
\caption{A sketch of an infalling reference frame in an 
 Eddington-Finkelstein diagram:\ the horizontal and vertical axes 
 are $r$ and $t^* = t+r^*-r$, respectively, where $r^*$ is the tortoise 
 coordinate.  The thick (blue) line denotes the spacetime trajectory 
 of the origin, $p_0$, of the reference frame, while the thin (red) 
 lines represent past-directed light rays emitted from $p_0$.  The 
 shaded area is the causal patch associated with the reference frame, 
 and the dotted (green) line represents the stretched ``horizon'' 
 as viewed from this reference frame.}
\label{fig:patch}
\end{figure}

In this infalling description, we expect that a ``horizon'' signaling 
the breakdown of the semiclassical description lies in the directions 
associated with ``past-directed and inward'' light rays (the directions 
with increasing $r$ and decreasing $t$ after $p_0$ crosses $r = 2 M 
l_{\rm P}^2$) as viewed from $p_0$; see Fig.~\ref{fig:patch}.%
\footnote{This ``horizon,'' as viewed from an infalling reference frame, 
 should not be confused with the stretched, or Schwarzschild, horizon 
 as viewed from a distant reference frame.}
As in the stretched horizon in a distant reference frame, this ``horizon'' 
emerges because of the ``squeezing'' of equal-time hypersurfaces; in 
particular, an observer following the trajectory of $p_0$ may probe 
only a tiny region near the Schwarzschild horizon for signals arising 
from this surface.  (Note that $-r$ plays a role of time inside 
the Schwarzschild horizon.)  Considering angular directions, this 
``horizon'' has an area of order $M^2 l_{\rm P}^4$, and can be 
regarded as being located at distances of order $M l_{\rm P}^2$ 
away from $p_0$ (with an appropriately defined distance measure 
on generic equal-time hypersurfaces in the infalling reference 
frame; see Section~\ref{subsec:consistency}).

In analogy with the case of a distant frame description, we denote 
basis states for the general microstates in an infalling reference 
frame (before $p_0$ reaches the singularity) as
\begin{equation}
  \ket{\Psi_{\bar{\alpha}\, \alpha\, \alpha_{\rm far}; \kappa}(M)},
\label{eq:states-falling}
\end{equation}
where $\bar{\alpha}$ labels the excitations of the ``horizon,'' and 
$\alpha$, and $\alpha_{\rm far}$ are the indices labeling the semiclassical 
excitations near and far from the black hole, conveniently defined; 
$\kappa$ is the vacuum index in an infalling reference frame, representing 
degrees of freedom that cannot be resolved by semiclassical operators.%
\footnote{After $p_0$ hits the singularity, the system as viewed from 
 the infalling reference frame can only be represented by ``singularity 
 states'':\ intrinsically quantum gravitational states that do not 
 allow for a spacetime interpretation~\cite{Nomura:2011rb}.}
The complementarity transformation provides a map from the basis states 
in a distant description, Eq.~(\ref{eq:states}), to those in an infalling 
description, Eq.~(\ref{eq:states-falling}), and vice versa.  The general 
form of this transformation can be quite complicated, depending, e.g., 
on equal-time hypersurfaces taken in the two descriptions (which are 
in turn related with the general procedure of erecting reference frames 
by standard coordinate transformations within each causal patch).  Here 
we consider how various indices are related under the transformation, 
focusing on the near black hole region.

Imagine that equal-time hypersurfaces in the two---distant and 
infalling---reference frames agree at some time $t=t_0$ in the spacetime 
region near but outside the surface where the stretched horizon exists 
if viewed from the distant reference frame.  (Note that the stretched 
horizon has physical substance only in a distant reference frame.) 
We are interested in how basis states in the two descriptions transform 
between each other in the timescale of the fall of the infalling reference 
frame.  The time here can be taken as the proper time at $p_0$ in each 
reference frame~\cite{Nomura:2011rb,Nomura:2013nya}, which is approximately 
the Schwarzschild time for the distant reference frame.  In this 
case, the relevant timescale is $t - t_0 \lesssim O(M l_{\rm P}^2 
\ln(M l_{\rm P}))$ in the distant reference frame, while $t - t_0 
\lesssim O(M l_{\rm P}^2)$ in the infalling reference frame.

As discussed in Section~\ref{subsec:obj}, in the distant reference frame, 
an object dropped from some $r_0$ with $r_0 - 2Ml_{\rm P}^2 \approx 
O(M l_{\rm P}^2)$ is first represented by $a$ and then by $\bar{a}$ 
after it hits the stretched horizon.  On the other hand, in the infalling 
frame, the object is represented by the index $\alpha$ throughout, 
first as a semiclassical excitation outside the Schwarzschild horizon 
and then as a semiclassical excitation inside the Schwarzschild horizon, 
implying that the object does not find anything special at the horizon. 
Here, we have assumed that $p_0$ follows (approximately) the trajectory 
of the falling object.  This suggests that a portion of the $\alpha$ 
index representing excitations in the interior of the black hole is 
transformed into the $\bar{a}$ index in the distant description (and 
vice versa) under the complementarity transformation; i.e., the interior 
of the black hole accessible from the infalling reference frame is 
encoded in the excitations of the stretched horizon in the distant 
reference frame.  Note that the amount of information needed to reconstruct 
the interior (in the semiclassical sense) is much smaller than the 
Bekenstein-Hawking entropy~\cite{'tHooft:1993gx,Nomura:2013lia}---the 
logarithm of the dimension of the relevant Hilbert space is of order 
$({\cal A}/l_{\rm P}^2)^q$ with $q < 1$.

In the exterior spacetime region, the portion of the $\alpha$ index 
representing excitations there, as well as the $\alpha_{\rm far}$ index, 
are mapped to the corresponding $a$ and $a_{\rm far}$ indices, and 
vice versa (after matching the equal-time hypersurface in the two 
descriptions through appropriate time evolutions).  Because equal-time 
hypersurfaces foliate the causal patch, excitations in the far exterior 
region naturally have trans-Planckian energies in the infalling 
description.  However, as discussed in Section~\ref{subsec:obj}, this 
does not mean that the semiclassical description is invalid---objects 
may still be described as excitations in the semiclassical spacetime, 
although stringy effects may become important.  Indeed, we expect that 
the semiclassical description is applicable in the far exterior region 
even in the infalling reference frame, because of the absence of the 
``squeezing'' effect described above which leads to the breakdown of 
the semiclassical picture.

We emphasize that the construction of the interior spacetime 
described here does not suffer from the paradoxes discussed in 
Refs.~\cite{Almheiri:2012rt,Almheiri:2013hfa,Marolf:2013dba}. 
By labeling states in terms of excitations, we are in a sense 
representing the interior spacetime already in the distant description. 
(The interpretation, however, is different.  In the distant description, 
the relevant excitations must be regarded as those of the stretched 
horizon.)  In fact, we do not find any inconsistency in postulating 
that the dynamics of an infalling object is described by the corresponding 
Hamiltonian in the semiclassical theory in a sufficiently small region 
around $p_0$, to the extent that microscopic details of interactions 
with $\kappa$ degrees of freedom are neglected.  Namely, we do not 
find any inconsistency in postulating that physics at the classical 
level is well described by general relativity.

Finally, we discuss where the fine-grained vacuum degrees of freedom 
represented by $\kappa$ must be viewed as being located in the infalling 
description.  Because of the lack of an obvious static limit, it is not 
straightforward to answer to this question.  Nevertheless, it seems 
natural to expect, in analogy with the case of a distant description, 
that most of the degrees of freedom are located close to the ``horizon'' 
(in terms of a natural distance measure in which the distance between 
the ``horizon'' and $p_0$ is of order $M l_{\rm P}^2$).  In fact, we 
expect that the number of $\kappa$ degrees of freedom existing around 
$p_0$ within a distance scale sufficiently smaller than $M l_{\rm P}^2$ 
is of $O(1)$ or smaller, since the time and length scales of the system 
characterizing local deviations from Minkowski space (as viewed from 
the infalling reference frame) are both of order $M l_{\rm P}^2$. 
As in the case of the distant description, we expect that the $\kappa$ 
degrees of freedom do not extend significantly to the far exterior 
region, since the existence of the black hole does not affect the 
spacetime there much.%
\footnote{Note that the descriptions in the two reference frames 
 are already different at the semiclassical level.  For example, the 
 backreaction of a detector click in a distant reference frame is 
 described as an absorption of a particle in the thermal bath, while 
 in an infalling reference frame it is described as an emission of 
 a particle, with the difference arising from different definitions 
 of energy in the two reference frames~\cite{Unruh:1983ms}.  The 
 reference frame dependence discussed here is much more drastic, 
 however---the spacetime locations of physical degrees of freedom 
 are different in the two reference frames.}

\subsection{Consistency between the distant and infalling descriptions}
\label{subsec:consistency}

In analyzing a black hole system in a distant reference frame, we argued 
that the microscopic information about the black hole, represented by the 
$k$ index, is distributed according to the gravitational thermal entropy 
calculated using semiclassical field theory.  In particular, on the 
Schwarzschild (or stretched) horizon, this information has a Planckian 
density:\ one qubit per area of order $l_{\rm P}^2$ on the horizon (or 
per volume of order $l_{\rm P}^3$ if we take into account the ``thickness'' 
of the stretched horizon, $\sim l_{\rm P}$).  On the other hand, we 
have just argued that in an infalling reference frame, the spacetime 
distribution of the microscopic information (now represented by the 
$\kappa$ index) is different.  In particular, the spatial density of 
the information around the Schwarzschild horizon, when the origin of 
the reference frame passes through it, is very small:\ one qubit per 
volume of order $(M l_{\rm P}^2)^3$.  How can we reconcile these two 
seemingly very different perspectives?

In this subsection, we consider this problem and argue that despite 
the fact that the spacetime distribution of the microscopic information 
depends on the reference frame one chooses to describe the system, 
the answers to any operationally well-defined question one obtains 
in different reference frames are consistent with each other.  As an 
example most relevant to our discussion, we consider a physical detector 
hovering at a constant Schwarzschild radius $r = r_{\rm d}$ ($> 2 M 
l_{\rm P}^2$).  In a distant description, the spatial density of the 
microscopic information, represented by $k$, is large at the location 
of the detector when $r_{\rm d} - 2Ml_{\rm P}^2 \ll M l_{\rm P}^2$. 
Such a detector (or a system of detectors) can thus be used for black 
hole mining:\ accelerated extraction of energy and information from 
the black hole.  In an infalling reference frame, however, the density 
of the microscopic information, represented by $\kappa$, is very small 
at the detector location, at least when the origin of the reference 
frame, $p_0$, passes nearby.  This implies that the rate of extracting 
information from spacetime cannot be much faster than $1/M l_{\rm P}^2$ 
around $p_0$ in the infalling description, reflecting the fact that 
the spacetime appears approximately flat there.  How are these two 
descriptions consistent?

In the distant description, the rate of extracting microscopic information 
about the black hole is at most of order one qubit per Schwarzschild 
time $1/T_{\rm H} = 8\pi M l_{\rm P}^2$ {\it per channel}, regardless 
of the location of the detector~\cite{Brown:2012un}---the acceleration 
of information extraction occurs not because of a higher speed of 
information extraction in each channel but because of an increased 
number of channels available by immersing the detector deep into the 
zone.  This implies that each single detector, which we define to act 
on a single channel, ``clicks'' once (i.e.\ extracts of $O(1)$ qubits) 
per a Schwarzschild time of order $8\pi M l_{\rm P}^2$.

Now, consider describing such a detector in an infalling reference 
frame whose origin $p_0$ is released at $r = 2 M l_{\rm P}^2 + 
O(M l_{\rm P}^2)$ from rest, at an angular location close to the 
detector.  To understand the relevant kinematics, we adopt the 
near-horizon Rindler approximation:\ for $r > 2 M l_{\rm P}^2$
\begin{equation}
  \rho \approx 2\sqrt{2 M l_{\rm P}^2 (r-2 M l_{\rm P}^2)},
\qquad
  \omega \approx \frac{t}{4 M l_{\rm P}^2},
\label{eq:Rinder-coord}
\end{equation}
in terms of which the metric is given by
\begin{equation}
  ds^2 \approx -\rho^2 d\omega^2 + d\rho^2 + r(\rho)^2 d\Omega.
\label{eq:Rinder-metric}
\end{equation}
As is well-known, this metric can be written in the Minkowski form
\begin{equation}
  ds^2 \approx -dT^2 + dZ^2 + r(T,Z)^2 d\Omega,
\label{eq:Mink-metric}
\end{equation}
by introducing the coordinates
\begin{equation}
  T = \rho \sinh\omega,
\qquad
  Z = \rho \cosh\omega,
\label{eq:Mink-coord}
\end{equation}
which can be extended into the $r < 2 M l_{\rm P}^2$ region.  Our setup 
corresponds to the situation in which the detector follows a trajectory 
of a constant $\rho$:
\begin{equation}
  \rho = \rho_{\rm d} \ll M l_{\rm P}^2,
\label{eq:detector}
\end{equation}
while the origin of the reference frame $p_0$---or the (fictitious) 
observer---is at a constant $Z$:
\begin{equation}
  Z = Z_{\rm o} \approx O(M l_{\rm P}^2).
\label{eq:p_0}
\end{equation}
Note that while we {\it approximate} the geometry by flat space, 
given by Eq.~(\ref{eq:Rinder-metric}) or (\ref{eq:Mink-metric}), 
the actual system has small nonzero curvature with lengthscale 
of order $M l_{\rm P}^2$.

As discussed above, the detector extracts an $O(1)$ amount of information 
in each time interval of
\begin{equation}
  \varDelta \omega \approx O\biggl( \frac{1}{4 M l_{\rm P}^2 T_H} \biggr) 
  \approx O(1),
\label{eq:del-omega}
\end{equation}
while the ``observer,'' $p_0$, and the detector meet (or pass by each 
other) at
\begin{equation}
  \left( \begin{array}{c} \omega \\ \rho \end{array} \right) 
  = \left( \begin{array}{c} {\rm arccosh}\frac{Z_{\rm o}}{\rho_{\rm d}} \\ 
    \rho_{\rm d} \end{array} \right) 
  \equiv \left( \begin{array}{c} \omega_* \\ \rho_* \end{array} \right).
\label{eq:meeting}
\end{equation}
This implies that in the Minkowski coordinates---i.e.\ as viewed from 
the infalling observer $p_0$---the detector clicks only once in each 
time/space interval of
\begin{align}
  \varDelta T &\approx \varDelta\omega \frac{\partial T}{\partial\omega} 
    \biggr|_{\omega=\omega_*, \rho=\rho_*} 
  \approx Z_{\rm o} \approx O(M l_{\rm P}^2),
\label{eq:click-T}\\
  \varDelta Z &\approx \varDelta\omega \frac{\partial Z}{\partial\omega} 
    \biggr|_{\omega=\omega_*, \rho=\rho_*} 
  \approx Z_{\rm o} \approx O(M l_{\rm P}^2),
\label{eq:click-Z}
\end{align}
around $p_0$.  This is precisely what we expect from the equivalence 
principle:\ the spacetime appears approximately flat when viewed 
from an infalling observer, up to curvature effects with lengthscale 
of $M l_{\rm P}^2$.  While the detector clicks of order $\ln(M l_{\rm P})$ 
times within the causal patch of the infalling reference frame, all 
these clicks occur at distances of order $M l_{\rm P}^2$ away from 
$p_0$, where we expect a higher density of $\kappa$ degrees of freedom. 
The two descriptions---distant and infalling---are therefore consistent, 
despite the fact that the spacetime distributions of the microscopic 
information about the black hole---represented by $k$ and $\kappa$, 
respectively---are different in the two reference frames.

While we have so far discussed the case in which a physical detector 
is located close to the Schwarzschild horizon, the conclusion is 
the same in the case of spontaneous Hawking emission.  In this case, 
since Hawking particles appear as semiclassical excitations only at 
$r - 2 M l_{\rm P}^2 \gtrsim M l_{\rm P}^2$ with local energies of 
order $1/M l_{\rm P}^2$, the consistency of the two descriptions is 
in a sense obvious.  Alternatively, one can regard this case as the 
$\rho_{\rm d} \approx M l_{\rm P}^2$ limit of the previous analysis. 
While the Rindler approximation is strictly valid only for $\rho$ 
sufficiently smaller than $M l_{\rm P}^2$, qualitative results are 
still valid for $\rho_{\rm d} \approx M l_{\rm P}^2$; in particular, 
the estimates in Eqs.~(\ref{eq:click-T},~\ref{eq:click-Z}) are valid 
at an order of magnitude level.

\subsection{Other reference frames---free fall from a nearby point}
\label{subsec:others}

In this subsection, we consider how the black hole is described in a class 
of reference frames whose origin follows a timelike geodesic released 
from rest at $r = r_0$, where $r_0$ is {\it close to} the Schwarzschild 
horizon, $r_0 - 2 M l_{\rm P}^2 \ll M l_{\rm P}^2$.%
\footnote{In a full geometry in which the black hole is formed by 
 collapsing matter, the trajectory of the origin, $p_0$, of such 
 a reference frame corresponds to a fine-tuned one in which $p_0$ 
 stays near outside of the Schwarzschild horizon for long time due 
 to large outward velocities at early times.  (Here, we have focused 
 only on the relevant branch in the full quantum state; see, e.g., 
 footnote~\ref{ft:stochastic}.)}
We argue that the description in these reference frames does not 
look similar to either the distant or infalling description discussed 
before, and yet it is consistent with both of them.%
\footnote{Note that we use the term ``infalling reference frame'' exclusively 
 for reference frames discussed in Sections~\ref{subsec:interior} and 
 \ref{subsec:consistency}, i.e.\ the ones in which $p_0$ starts from 
 rest at $r_0$ with $r_0 - 2 M l_{\rm P}^2 \gtrsim O(M l_{\rm P}^2)$.}

To understand how the black hole appears in such a reference frame, let us 
consider a setup similar to that in Section~\ref{subsec:consistency}---a 
physical detector hovering at a constant Schwarzschild radius $r = 
r_{\rm d}$---and see how this system is described in the reference 
frame.  As in Section~\ref{subsec:consistency}, we may adopt the Rindler 
approximation, in which Eq.~(\ref{eq:p_0}) is now replaced by
\begin{equation}
  Z = Z_{\rm o} \ll M l_{\rm P}^2.
\label{eq:p_0-2}
\end{equation}
This implies that as viewed from this reference frame, the detector 
clicks once in each time/space interval of
\begin{equation}
  \varDelta T \approx \varDelta Z \approx Z_{\rm o} \ll M l_{\rm P}^2.
\label{eq:click-TZ}
\end{equation}
Here, we have assumed that $\rho_{\rm d} < Z_{\rm o}$.  Since each 
detector click extracts an $O(1)$ amount of information from spacetime, 
which we expect not to occur in Minkowski space, this implies that the 
spacetime cannot be viewed as approximately Minkowski space over a 
region beyond lengthscale $Z_{\rm o}$.  In particular, in contrast 
with the case in an infalling reference frame (with $Z_{\rm o} \gtrsim 
O(M l_{\rm P}^2)$), the spacetime region around $p_0$ in this reference 
frame does not appear nearly flat over lengthscale of $M l_{\rm P}^2$ 
when $p_0$ crosses the Schwarzschild horizon.

At a technical level, this difference arises from the fact that the 
relative boost of $p_0$ with respect to the distant reference frame 
when $p_0$ approaches the detector
\begin{equation}
  \gamma = \frac{1}{\sqrt{1-v_{\rm rel}^2}} 
  = \sqrt{\frac{ 1 - \frac{2 M l_{\rm P}^2}{r_0} }
      { 1 - \frac{2 M l_{\rm P}^2}{r_d} }},
\label{eq:rel-vel}
\end{equation}
is very different in the two reference frames.  In an infalling 
reference frame $\gamma$ is huge, $\approx O(M l_{\rm P}^2/\rho_{\rm d})$, 
while in the reference frame considered here $\gamma \approx 
O(Z_{\rm o}/\rho_{\rm d})$, which is not as large as that 
in the infalling case.  In the infalling reference frame of 
Sections~\ref{subsec:interior} and \ref{subsec:consistency}, 
the huge boost of $\gamma \approx O(M l_{\rm P}^2/\rho_{\rm d})$ 
``stretched'' the interval between detector clicks to time/length 
scales of order $M l_{\rm P}^2$.  Here, this ``stretching'' makes 
only a small region around $p_0$, with lengthscale of order $Z_{\rm o}$ 
($\ll M l_{\rm P}^2$), look nearly flat at any given time.

We may interpret this result to mean that in the reference frame 
under consideration, the ``horizon'' (as viewed from this reference 
frame) is located at a distance of order $Z_{\rm o}$ away from $p_0$, 
so that detector clicks occur near or ``on'' this surface.  (In the 
latter case, the detector click events must be viewed as occurring 
in the regime outside the applicability of the semiclassical description; 
in particular, they can only be described as complicated quantum 
gravitational processes occurring on the ``horizon.'')  Since we 
expect that microscopic information about the black hole (analogous 
to $k$ and $\kappa$ in the distant and infalling reference frames, 
respectively) is located near and on the ``horizon,'' there is no 
inconsistency that detector clicks extract microscopic information 
from the black hole.

One might be bothered by the fact that in this reference frame 
spacetime near the Schwarzschild horizon does not appear large, 
$\approx O(M l_{\rm P}^2)$, nearly flat space, and consider that 
this implies the non-existence of a large black hole interior as 
suggested by general relativity.  This is, however, not correct. 
The existence of {\it a} reference frame in which spacetime around 
the Schwarzschild horizon appears as a large nearly flat region---in 
particular, the existence of an infalling reference frame discussed 
in Sections~\ref{subsec:interior} and \ref{subsec:consistency}---already 
ensures that an infalling physical object/observer does not experience 
anything special, e.g.\ firewalls, when it/he/she crosses the 
Schwarzschild horizon.  The analysis given here simply says that 
the spacetime around the Schwarzschild horizon does {\it not always} 
appear as a large nearly flat region, even in a reference frame whose 
origin falls freely into the black hole.  This extreme relativeness 
of descriptions is what we expect from complementarity.

\subsection{(Non-)relations with the Unruh effect in Minkowski space}
\label{subsec:Unruh}

It is often thought that the system described above is similar to 
an accelerating detector existing in Minkowski space, based on a 
similarity of geometries between the two setups.  If this were true 
at the full quantum level, it would mean that the description in 
an {\it inertial} reference frame in Minkowski space must possess a 
``horizon,'' at which the semiclassical description of the system 
breaks down.  Does this make sense?

Here we argue that physics of a detector held near the Schwarzschild 
horizon, given above in Section~\ref{subsec:others}, is, in fact, 
different from that of an accelerating detector in Minkowski space. 
The intuition that the two must be similar comes from the (wrong) 
perception that the detector located near the Schwarzschild horizon 
feels a high blueshifted Hawking temperature, $\approx 1/\rho_{\rm d} 
\gg 1/M l_{\rm P}^2$, which makes the detector click at a high rate, 
while the spacetime curvature there is very small, with lengthscale 
$\approx M l_{\rm P}^2$, so that such a tiny curvature must not affect 
the system.  This intuition, however, is flawed by mixing up two 
different pictures---the system as viewed at the location of the 
detector and as viewed in the asymptotic region.

Suppose we represent all quantities as defined in the asymptotic 
region.  The temperature a detector feels is then of order $1/M 
l_{\rm P}^2$ and the timescale for detector clicks is $T \approx 
O(M l_{\rm P}^2)$ for {\it any} $r_{\rm d} > 2 M l_{\rm P}^2$.  On 
the other hand, the energy density of the black hole region is of 
order $M/(M l_{\rm P}^2)^3$, so that the curvature lengthscale $L$ 
is estimated as
\begin{equation}
  \frac{1}{L^2} \sim G_{\rm N} \frac{M}{(M l_{\rm P}^2)^3} 
  \sim \frac{1}{(M l_{\rm P}^2)^2}.
\label{eq:curv-1}
\end{equation}
This implies that
\begin{equation}
  T \sim L \sim O(M l_{\rm P}^2);
\label{eq:TL-1}
\end{equation}
namely, curvature is expected to give {\it an $O(1)$ effect} on the 
dynamics of the detector response.

The same conclusion can also be reached when we represent all the 
quantities in the static frame at the detector location.  In this 
case, the temperature the detector feels is of order $1/M l_{\rm P}^2 
\chi$, where $\chi = \sqrt{1 - 2 M l_{\rm P}^2/r_{\rm d}}$ is the 
redshift factor, so that $T \approx O(M l_{\rm P}^2 \chi)$.  On the 
other hand, the energy density of the black hole region is given by 
$\sim (M/\chi)/(M l_{\rm P}^2)^3 \chi$, so that the ``blueshifted 
curvature length'' $L$ is given by
\begin{equation}
  \frac{1}{L^2} \sim G_{\rm N} \frac{M/\chi}{(M l_{\rm P}^2)^3 \chi} 
  \sim \frac{1}{(M l_{\rm P}^2 \chi)^2}.
\label{eq:curv-2}
\end{equation}
This yields
\begin{equation}
  T \sim L \sim O(M l_{\rm P}^2 \chi),
\label{eq:TL-2}
\end{equation}
again implying that curvature provides an $O(1)$ effect on the dynamics.

It is, therefore, no surprise that the physics of a near-horizon 
detector in Section~\ref{subsec:others} differs significantly from 
that of an accelerating detector in Minkowski space experiencing 
the Unruh effect~\cite{Unruh:1976db}.  In fact, we consider, as we 
naturally expect, that an inertial frame description in Minkowski 
space does {\it not} have a horizon, implying that no information 
about spacetime is extracted by an accelerating detector, despite 
the fact that it clicks at a rate controlled by the acceleration 
$a$, $T \approx O(1/a)$, in the detector's own frame.  This is indeed 
consistent with the idea that any information must be accompanied 
by energy.  In the black hole case, the detector mines the black hole, 
i.e.\ its click extracts energy from the black hole spacetime, while 
in the Minkowski case the energy needed to excite the detector comes 
entirely from the force responsible for the acceleration of the 
detector---the detector does not mine energy from Minkowski space. 
We conclude that blueshifted Hawking radiation and Unruh radiation 
in Minkowski space are very different as far as the information 
flow is concerned.

Does this imply a violation of the equivalence principle?  The 
equivalence principle states that gravity is the same as acceleration, 
and the above statement might seem to contradict this principle.  This 
is, however, not true.  The principle demands the equivalence of the 
two only at a point in space in a given coordinate system, and the 
descriptions of the two systems discussed above---a black hole and 
Minkowski space---are indeed the same in an infinitesimally small 
(or lengthscale of order $l_*$) neighborhood of $p_0$.  The principle 
does not require that the descriptions must be similar in regions 
away from $p_0$, and indeed they are very different:\ there is a 
``horizon'' at a distance of order $Z_{\rm o}$ from $p_0$ in the 
black hole case while there is no such thing in the Minkowski case. 
And it is precisely in these regions that the detector clicks to 
extract (or non-extract) information from the black hole (Minkowski) 
spacetime.  In quantum mechanics, a system is specified by a quantum 
state which generally encodes global information on the equal-time 
hypersurface.  It is, therefore, natural that the equivalence principle, 
which makes a statement only about a point, does not enforce the 
equivalence between physics of blueshifted Hawking radiation and 
of the Unruh effect in Minkowski space at the fully quantum level.

\subsection{Complementarity:\ general covariance in quantum gravity}
\label{subsec:compl}

We have argued that unitary information transfer described in 
Section~\ref{sec:distant}, associated with Hawking emission and 
black hole mining, is consistent with the existence of the interior 
spacetime suggested by general relativity.  We can summarize important 
lessons we have learned about quantum gravity through this study 
in the following three points:
\begin{itemize}
\item
In {\it a fixed reference frame}, the microscopic information about 
spacetime, in this case about a black hole, may be viewed as being 
associated with specific spacetime locations.  In particular, for 
a (quasi-)static description of a system, these degrees of freedom 
are distributed according to the gravitational thermal entropy calculated 
using semiclassical field theory.  The distribution of these degrees 
of freedom---which we may call ``constituents of spacetime''---controls 
how they can interact with the degrees of freedom in semiclassical 
theory, e.g.\ matter and radiation in semiclassical field theory.
\item
The spacetime distribution of the microscopic information, however, 
changes if we adopt a different reference frame to describe the 
system.  In this sense, the ``constituents of spacetime'' are 
{\it not} anchored to spacetime; they are associated with specific 
spacetime locations only after the reference frame is fixed. 
In particular, no reference frame independent statement can 
be made about where these degrees of freedom are located in 
spacetime.  We may view this as a manifestation of the holographic 
principle~\cite{'tHooft:1993gx,Susskind:1994vu}---gauge invariant 
degrees of freedom in a quantum theory of gravity live in some 
``holographic space.''
\item
Despite the strong reference frame dependence of the location of the 
microscopic degrees of freedom, the answers to any physical question 
are consistent with each other when asked in different reference frames. 
In particular, when we change the reference frame, the distribution of 
the microscopic degrees of freedom (as well as some of the semiclassical 
degrees of freedom) is rearranged such that this consistency is 
maintained.
\end{itemize}
These items are basic features of general coordinate transformations 
at the level of full quantum gravity, beyond the approximation 
of semiclassical theory in curved spacetime.  In particular, they 
provide important clues about how complementarity as envisioned 
in Refs.~\cite{Nomura:2011rb,Nomura:2013nya} may be realized at 
the microscopic level.

\section{Summary---A Grand Picture}
\label{sec:summary}

The relation between the quantum mechanical view of the world and 
the spacetime picture of general relativity has never been clear. 
The issue becomes particularly prominent in a system with a black 
hole.  Quantum mechanics suggests that the black hole formation and 
evaporation processes are unitary---a black hole appears simply as 
an intermediate (gigantic) resonance between the initial collapsing 
matter and final Hawking radiation states.  On the other hand, general 
relativity suggests that a classical observer falling into a large 
black hole does not feel anything special at the horizon.  These two, 
seemingly unrelated, assertions are surprisingly hard to reconcile. 
With naive applications of standard quantum field theory on curved 
spacetime, one is led to the conclusion that unitarity of quantum 
mechanics is violated~\cite{Hawking:1976ra} or that an infalling 
observer finds something dramatic (firewalls) at the location of 
the horizon~\cite{Almheiri:2012rt,Almheiri:2013hfa,Marolf:2013dba,%
Braunstein:2009my}.

In this paper, we have argued that the resolution to this puzzle 
lies in how a semiclassical description of the system---quantum 
theory of matter and radiation on a fixed spacetime background---arises 
from the microscopic theory of quantum gravity.  While a semiclassical 
description employs an {\it exact} spacetime background, the quantum 
uncertainty principle implies that there is no such thing---there 
is an intrinsic uncertainty for background spacetime for any finite 
energy and momentum.  This implies, in particular, that at the microscopic 
level there are many different ways to arrive at the same background 
for the semiclassical theory, within the precision allowed by quantum 
mechanics.  This is the origin of the Bekenstein-Hawking (and related, 
e.g.\ Gibbons-Hawking~\cite{Gibbons:1977mu}) entropy.  The semiclassical 
picture is obtained after coarse-graining these degrees of freedom 
representing the microscopic structure of spacetime, which we called 
the vacuum degrees of freedom.  More specifically, any result in 
semiclassical theory is a statement about the maximally mixed ensemble 
of microscopic quantum states consistent with the specified background 
within the required uncertainty~\cite{Nomura:2014yka}.

This picture elucidates why the purely semiclassical calculation of 
Ref.~\cite{Hawking:1976ra} finds a violation of unitarity.  At the 
microscopic level, formation and evaporation of a black hole are 
processes in which information in the initial collapsing matter is 
converted into that in the vacuum degrees of freedom, which is later 
transferred back to semiclassical degrees of freedom, i.e.\ Hawking 
radiation.  Since semiclassical theory is incapable of describing 
microscopic details of the vacuum degrees of freedom (because it 
describes them as already coarse-grained, Bekenstein-Hawking entropy), 
the {\it description} of the black hole formation and evaporation 
processes in semiclassical theory violates unitarity at all stages 
throughout these processes.  This, of course, does not mean that 
the processes are non-unitary at the fundamental level.

In order to address the unitary evolution and explore its relation with 
the existence or non-existence of the interior spacetime, we therefore 
need to discuss the properties of the vacuum degrees of freedom.  While 
the theory governing the detailed microscopic dynamics of these degrees 
of freedom is not yet fully known, we may include them in our description 
in the form of a new index---vacuum index---carried by the microscopic 
quantum states (which we denoted by $k$ and $\kappa$) in addition to 
the indices representing excitations in semiclassical theory and of the 
stretched horizon.  We have argued that these degrees of freedom show 
peculiar features, which play key roles in addressing the paradoxes 
discussed in Refs.~\cite{Almheiri:2012rt,Almheiri:2013hfa,Marolf:2013dba}:

\vspace{4mm}
{\bf Extreme relativeness:}
\begin{itemize}
\item[]
In a fixed reference frame, vacuum degrees of freedom may be viewed 
as distributed (nonlocally) over space.  The spacetime distribution 
of these degrees of freedom, however, changes if we adopt a different 
reference frame---they are not anchored to spacetime, and rather 
live in some ``holographic space.''  This dependence on the reference 
frame occurs in a way that the answers to any physical question 
are consistent with each other when asked in different reference 
frames. Together with the reference frame dependence of (some of 
the) semiclassical degrees of freedom, discussed in the earlier 
literature~\cite{Susskind:1993if,Susskind:2005js}, this comprises 
basic features of how general coordinate transformations work in 
the full theory of quantum gravity.
\end{itemize}

\vspace{2mm}
{\bf Spacetime-matter duality:}
\begin{itemize}
\item[]
The vacuum degrees of freedom exhibit dual properties of spacetime 
and matter (even in a description in a single reference frame):\ while 
these degrees of freedom are interpreted as representing the way the 
semiclassical spacetime is realized at the microscopic level, their 
interactions with semiclassical degrees of freedom make them look like 
thermal radiation.  (At a technical level, the Hilbert space labeled 
by the vacuum index and that by semiclassical excitations do not 
factor.)  In a sense, these degrees of freedom are neither spacetime 
nor matter/radiation, as can be seen from the fact that their spacetime 
distribution changes as we change the reference frame, and that their 
detailed dynamics cannot be treated in semiclassical theory (as was 
done in Refs.~\cite{Almheiri:2012rt,Almheiri:2013hfa,Marolf:2013dba}). 
This situation reminds us of wave-particle duality, which played 
an important role in early days in the development of quantum 
mechanics---a quantum object exhibited dual properties of waves 
and particles, while the ``true'' (quantum) description did not 
fundamentally rely on either of these classical concepts.
\end{itemize}
\vspace{3mm}
These features make the existence of the black hole interior consistent 
with unitary evolution, in the sense of complementarity~\cite{Susskind:1993if} 
as envisioned in Refs.~\cite{Nomura:2011rb,Nomura:2013nya}.  In particular, 
a large nearly flat spacetime region near the Schwarzschild horizon 
becomes manifest in a reference frame whose origin follows a free-fall 
trajectory starting from rest from a point sufficiently far from the 
black hole.

It is often assumed that two systems related by the equivalence 
principle, e.g.\ a static detector held near the Schwarzschild horizon 
and an accelerating detector in Minkowski space, must reveal similar 
physics.  This is, however, not true.  Since the equivalence principle 
can make a statement only about a point at a given moment in a given 
reference frame, while a system in quantum mechanics is specified by 
a state which generally encodes global information on the equal-time 
hypersurface, there is no reason that physics of the two systems must 
be similar beyond a point in space.  In particular, a detector reacts 
very differently to blueshifted Hawking radiation and Unruh radiation 
in Minkowski space at the microscopic level---it extracts microscopic 
information about spacetime in the former case, while it does not 
in the latter.

While our study has focused on a system with a black hole, we do not 
see any reason why the basic picture we arrived at does not apply 
to more general cases.  We find it enlightening that our results 
indicate specific properties for the microscopic degrees of freedom 
that play a crucial role in the emergence of spacetime at the fundamental 
level.  Unraveling the detailed dynamics of these degrees of freedom 
would be a major step toward obtaining a complete theory of quantum 
gravity.  As a first step, it seems interesting to study implications 
of our picture for the case that spacetime approaches anti-de~Sitter 
space in the asymptotic region, in which we seem to know a little 
more~\cite{Maldacena:1997re}.  It would also be interesting to 
explore implications of our picture for cosmology, e.g.\ along the 
lines of Refs.~\cite{Nomura:2011rb,Nomura:2011dt,Nomura:2012zb}.

\section*{Acknowledgments}

We would like to thank Raphael Bousso, Ben Freivogel, Daniel Harlow, 
Juan Maldacena, Donald Marolf, Joseph Polchinski, Douglas Stanford, 
Jaime Varela, Erik Verlinde, Herman Verlinde, and I-Sheng Yang for 
various conversations during our exploration of this subject.  Y.N. 
thanks the Aspen Center for Physics and the National Science Foundation 
(NSF) Grant \#1066293 for hospitality during his visit in which a part 
of this work was carried out.  F.S. thanks the Department of Energy 
(DOE) National Nuclear Security Administration Stewardship Science 
Graduate Fellowship for financial support.  This work was supported 
in part by the Director, Office of Science, Office of High Energy and 
Nuclear Physics, of the U.S.\ DOE under Contract DE-AC02-05CH11231, 
and in part by the NSF under grant PHY-1214644.

\end{document}